\documentclass[journal,twoside, print]{ieeecolor}

\usepackage{cite}
\usepackage{amsmath,amssymb,amsfonts}
\usepackage{algorithmic}
\usepackage{graphicx}
\usepackage{textcomp}

\usepackage{listings}
\usepackage{booktabs}
\usepackage{arydshln} 
\usepackage{url}
\usepackage{hyperref}

\definecolor{subsectioncolor}{rgb}{0.0,0.0,0.0}
\definecolor{nblue}{cmyk}{0,0,0,1}
\definecolor{mblue}{cmyk}{0,0,0,1}

\def\BibTeX{{\rm B\kern-.05em{\sc i\kern-.025em b}\kern-.08em
    T\kern-.1667em\lower.7ex\hbox{E}\kern-.125emX}}

\begin{document}
\title{\LARGE Multi-Resolution 3D Convolutional Neural Networks for Automatic Coronary Centerline Extraction in Cardiac CT Angiography Scans}
\author{Zohaib Salahuddin, Matthias Lenga, Hannes Nickisch  \vspace{-1.3cm} 
\thanks{Zohaib Salahuddin is a student of Erasmus Mundus Joint Master Degree in Medical Imaging and Applications (MAIA) (e-mail: zohaib.salahuddin@studentmail.unicas.it).}
}

\maketitle
\pagestyle{empty}

\definecolor{codegreen}{rgb}{0,0.4,0}
\definecolor{codegray}{rgb}{0.5,0.5,0.5}
\definecolor{codepurple}{rgb}{0.58,0,0.82}
\lstdefinestyle{listingstyle}{
    backgroundcolor=\color{white},   
    commentstyle=\color{codegreen},
    keywordstyle=\color{red},
    numberstyle=\tiny\color{codegray},
    stringstyle=\color{codepurple},
    basicstyle=\ttfamily\footnotesize,
    breakatwhitespace=false,         
    breaklines=true,                 
    captionpos=b,                    
    keepspaces=true,
    numbers=left,                    
    numbersep=5pt,                  
    showspaces=false,                
    showstringspaces=false,
    showtabs=false,                  
    tabsize=2
}
\lstset{style=listingstyle}
\lstset{language=Python}

\begin{abstract}
We propose a deep learning-based automatic coronary artery tree centerline tracker (AuCoTrack) \cite[§13, p. 207]{salahuddin20aucotrack} extending the vessel tracker by \cite{WOLTERINK201946}. A dual pathway Convolutional Neural Network (CNN) operating on multi-scale 3D inputs predicts the direction of the coronary arteries as well as the presence of a bifurcation. A similar multi-scale dual pathway 3D CNN is trained to identify coronary artery endpoints for terminating the tracking process.
Two or more continuation directions are derived based on the bifurcation detection. The iterative tracker detects the entire left and right coronary artery trees based on only two ostium landmarks derived from a model-based segmentation of the heart \cite{Ecabert2008}. 

The 3D CNNs were trained on a proprietary dataset consisting of 43 CCTA scans. An average sensitivity of 87.1\% and clinically relevant overlap of 89.1\% was obtained relative to a refined manual segmentation. In addition, the MICCAI 2008 Coronary Artery Tracking Challenge (CAT08) training and test datasets were used to benchmark the algorithm and to assess its generalization. An average overlap of 93.6\% and a clinically relevant overlap of 96.4\% were obtained. The proposed method achieved better overlap scores than the current state-of-the-art automatic centerline extraction techniques on the CAT08 dataset with a vessel detection rate of 95\%.
\end{abstract}

\begin{IEEEkeywords}
coronary artery centerline tracking, coronary computed tomography angiography, convolutional neural networks
\end{IEEEkeywords}

\section{Introduction}
\label{sec:introduction}
Coronary artery disease is one of the leading causes of deaths worldwide. It was responsible for 9.43 million deaths in 2016 \cite{WHO2018}. Coronary arteries are responsible for supplying oxygenated blood to the heart muscles. Two main arteries branch off the aorta namely Left Main Coronary Artery (LCA) and Right Coronary Artery (RCA) which supply blood to the left and right parts of the heart, respectively. These two main arteries then divide into a network of smaller coronary arteries which wrap themselves around the heart. Coronary artery disease  (CAD) is the narrowing or blockage of one or several coronary arteries due to the build-up of cholesterol and fatty deposits called plaque on the inner lining of the arterial wall. This constriction can result in an inadequate supply of blood to the heart muscle which can be fatal \cite{Malakar2016}. Hence, there is a need for timely diagnosis and detection of this constriction in the arteries.

Coronary Angiography (CA) in the CathLab is an interventional procedure for coronary artery disease evaluation and treatment which provides information related to the coronary lumen. Coronary angiography requires a contrast agent and coronary catheterization as treatment can be performed during the intervention. Complications due to the invasive nature of coronary angiography occur in less than 2\% of the cases, with mortality of less than 0.08\% \cite{Tavakol2012}. Hence, there is a non-negligible risk associated with coronary angiography.
Computed Tomography Angiography (CCTA) is a non-invasive alternative which provides information on the extent and type of plaque present \cite{Peach201111}. CCTA  images have a high spatial resolution consisting of hundreds of slices. Manual reading of volumetric CCTA images is a time-consuming task even for trained experts due to the size and diversity of the arteries. Due to increasing number of CCTA scans, automatic analysis of CCTA images and improved 3D visualization is desirable. The radiation dose of CCTA is often less than the dose incurred by CA, in particular if the intervention involves revascularisation e.g. stenting.

There are many techniques to visualize the coronary arteries in CCTA images such as maximum intensity projection (MIP), volume rendering techniques (VRT), multiplanar reformatting (MPR) and curved multiplanar reformatting (cMPR) \cite{Cademartiri2007129}. Such advanced visualization techniques facilitate image reading and are, for example, used to guide stenosis and plaque detection \cite{Stimpel2018} and quantification. The computation of MPRs and cMPRs typically relies on centerlines of the coronary arteries. Hence, an important building block in the diagnosis of coronary artery disease is the extraction of coronary artery centerlines.

Manual extraction of coronary artery centerlines is time-consuming, error-prone and has a large inter-operator variability. In order to support the radiographer in the extraction of coronary artery centerlines, many interactive, semi-automatic and automatic methods for coronary centerline extraction have been proposed. The reformatted images obtained using centerlines can also be used for other purposes such as lumen segmentation of coronary arteries \cite{Huang2018608}.
Deep learning and machine learning-based methods typically use coronary artery centerline extraction as a preprocessing step for plaque identification and stenosis analysis \cite{Hampe20196}. A recurrent neural network was used by \cite{Zreik2019} to detect stenoses from multiplanar reformatted (MPR) images which were reconstructed using extracted coronary centerlines. Hence, an automatic coronary artery centerline extraction algorithm which provides consistent performance on CCTA images with variable image quality and calcium scores in a few seconds is desirable.

We propose a fully automatic coronary centerline extraction pipeline (called \textbf{AuCoTrack}) \cite[§13, p. 207]{salahuddin20aucotrack} based on a dual pathway multiscale 3D convolutional neural network. This pipeline contains three modules: The first module called Direction and Bifurcation Classification network (\textbf{DBC-Net}), is a multiscale 3D CNN for a local patch to determine the direction towards the center of the coronary artery with respect to the center of the patch as well as the patch type (normal or bifurcation). The second module namely Stop Patch Classification network (\textbf{STC-Net}), consists of another multiscale 3D CNN to determine if the patch contains the artery or not. The third module called \textbf{Tracker}, orchestrates the centerline extraction. The tracking is initialized at two ostium points obtained automatically. The tracker obtains predictions for directions and patch type from the DBC-Net for each patch. The tracker then takes steps in order to determine the centerline of the arteries. The tracker terminates based on the output of the STC-Net.

We evaluate \textbf{AuCoTrack} using four-fold cross validation on a proprietary dataset. In order to compare this approach to other state-of-the-art methods, an evaluation was conducted on the training and test dataset of the MICCAI Coronary Artery Tracking challenge 2008 (CAT08). Additionally, an analysis of the algorithm was performed to correlate the qualitative and quantitative analysis as well as the failure cases with research findings.

\begin{figure}[h!]
\centerline{\includegraphics[width =\columnwidth]{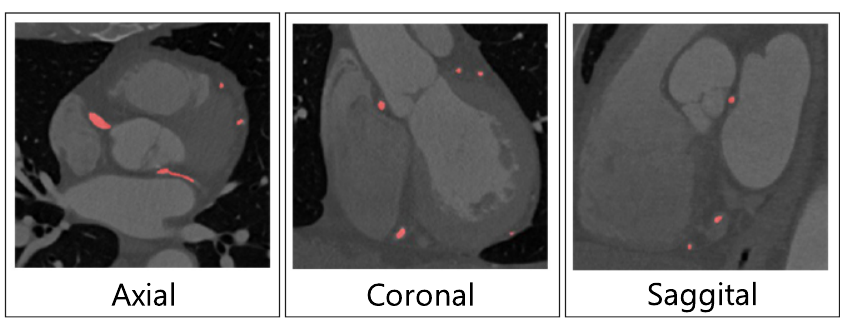}}
\caption{Axial, coronal and sagittal slices of a CCTA image. The coronary arteries are overlayed as red. The presence of vessel like structures around the heart make the extraction of coronary centerlines complicated.}
\end{figure}

\subsection{State of the art}
\label{sec:stateoftheart}
In order to extract the coronary artery centerlines, three types of approaches can be adopted: \textit{automatic}, \textit{semi-automatic} and \textit{interactive}. According to the MICCAI 2008 Coronary Artery Centerline Extraction Challenge (CAT08) guidelines as specified in \cite{Schaap2009701}, an approach may be defined as fully automatic if it utilizes no manually placed initialization points to track the entire coronary tree. If one point per vessel is provided to extract the coronary tree, the approach is said to be semi-automatic. If more than two points per vessel are required to obtain the coronary tree, it is labeled as an interactive approach. Extraction of the entire coronary tree based on interactive and semi-automatic approaches requires anatomical knowledge and manual inspection
 of the CCTA image to place points for each coronary artery individually. Since there are a lot of coronary arteries present in each coronary tree, these approaches increase the processing time. Hence, there is a need to establish a robust and automatic coronary artery centerline tracking algorithm which requires minimal user interaction.

A multiple hypothesis tracking approach based on a mathematical template vessel model combined with standard minimal paths method was used by \cite{Friman2020} to extract coronary artery centerlines. Standard minimal path-based methods experience shortcut issues and may require a lot of interaction to extract the entire vessel tree. At the time of this publication, 
\cite{Friman2020}'s method ranked first on MICCAI 2008 Coronary Artery Centerline Extraction Challenge (CAT08) as an interactive method. It requires 2.6 points on average per vessel and takes 6 min to extract four coronary arteries per CCTA image. \cite{Schaap2009701} used multivariate linear regression on image intensities to estimate an initial vessel boundary followed by a subsequent refinement of this result using non-linear regression. This method requires 2.2 points per vessel and takes 22 min to extract four coronary arteries per CCTA image. High processing times along with repeated user interaction per vessel is not desirable in clinical practice.

In \cite{Krissian2018}, morphological operations and denoising filters were used to obtain a region of interest. The probability of belonging to the coronary artery class for each voxel was then determined using a fuzzy classifier. The start points were automatically determined and the end points were provided manually for each vessel. A minimal path between these two points was traced based on voxel probability map generated by the classifier to obtain the centerlines. This semi-automatic method takes 7 h to extract four coronary arteries per CCTA image. \cite{Cetin2012} used a second order tensor constructed from
directional intensity measurements to track the entire coronary tree from a single seed point placed at the center of the cross-section of one of the vessels. This method utilizes an automatic branch detection based on K-means clustering of the intensity values. As a pre-processing step, a calcification filter is applied which requires annotations by an expert on the training CCTA scans. This methods takes 8 to 10 min on a 2.67 GHz dual processor to detect coronary arteries per CCTA scan. \cite{Cetin2015} also proposed an extension of this method to utilize  cylindrical flux-based higher order tensor (HOT) in 4D which also solves the problem of branch detection. This method takes 30 s to detect coronary arteries per CCTA scan on a computer equipped with an Intel Xeon X560 @ 2.67 GHz CPU and 64 GB memory.

State-of-the-art performance for automatic coronary centerline extraction was achieved by \cite{Zheng2013}. This method utilizes a segmentation mask to define a vessel specific region of interest (ROI) in order to constrain the centerline refinement by their model driven algorithm for extracting the main branches. The side branches are then traced by using region growing based on lumen segmentation. It was trained on 108 images of their proprietary dataset and takes 60 s to extract coronary arteries per CCTA scan. \cite{Kitamura2012} constructed a shape model of the coronary vessels and an Adaboost classifier in order to differentiate between normal and abnormal vessels for automatic centerline extraction. This method was trained on a proprietary dataset and the entire coronary tree centerline extraction takes 160 s per CCTA scan. \cite{Frangi1998} introduced a multiscale vessel enhancement filtering which obtained a vesselness measure based on eigenvalues of a Hessian. \cite{Yang2011} employed an improved version of Frangi's multiscale vessel enhancement filtering to obtain an initial tree which was further refined by branch searching automatically. This method takes 120 s on a standard desktop computer to track the entire coronary tree in a CCTA image. These runtime estimations have to be discounted by the recent advances in computational hardware.

Some methods utilize various handcrafted features such as virtual contrast and morphological operations. These handcrafted features are based on certain assumptions and they require explicit modeling in cases when the underlying assumptions do not hold e.g. bifurcations \cite{Frangi2000,Krissian2018,Wang2008,Cetin2012,Cetin2015}. 

Recently, an iterative CNN tracker was proposed in order to extract centerlines \cite{WOLTERINK201946}. This method does not require any handcrafted features. They propose a serial tracker that utilizes the direction and step-size predicted by the CNN in order to obtain the centerlines. They were able to achieve near state-of-the-art performance as compared to interactive methods. This method requires at least one seed point per vessel in order to extract its centerline. Some vessels require more than one point due to premature termination of the tracker. An additional CNN to extract seed points for the vessels was also proposed in order to make the algorithm automatic. However, the seed identification CNN requires training images in which all the coronary arteries have been annotated properly \cite{WOLTERINK201946}. Hence, this method requires 10 s to extract 4 coronary arteries per CCTA scan. Bifurcation detection in coronary arteries is a challenging task. \cite{WOLTERINK201946}'s CNN tracker extracts the coronary arteries in two directions without taking bifurcations into account.

\begin{figure*}[t!]
\centerline{\includegraphics[width=0.75\textwidth]{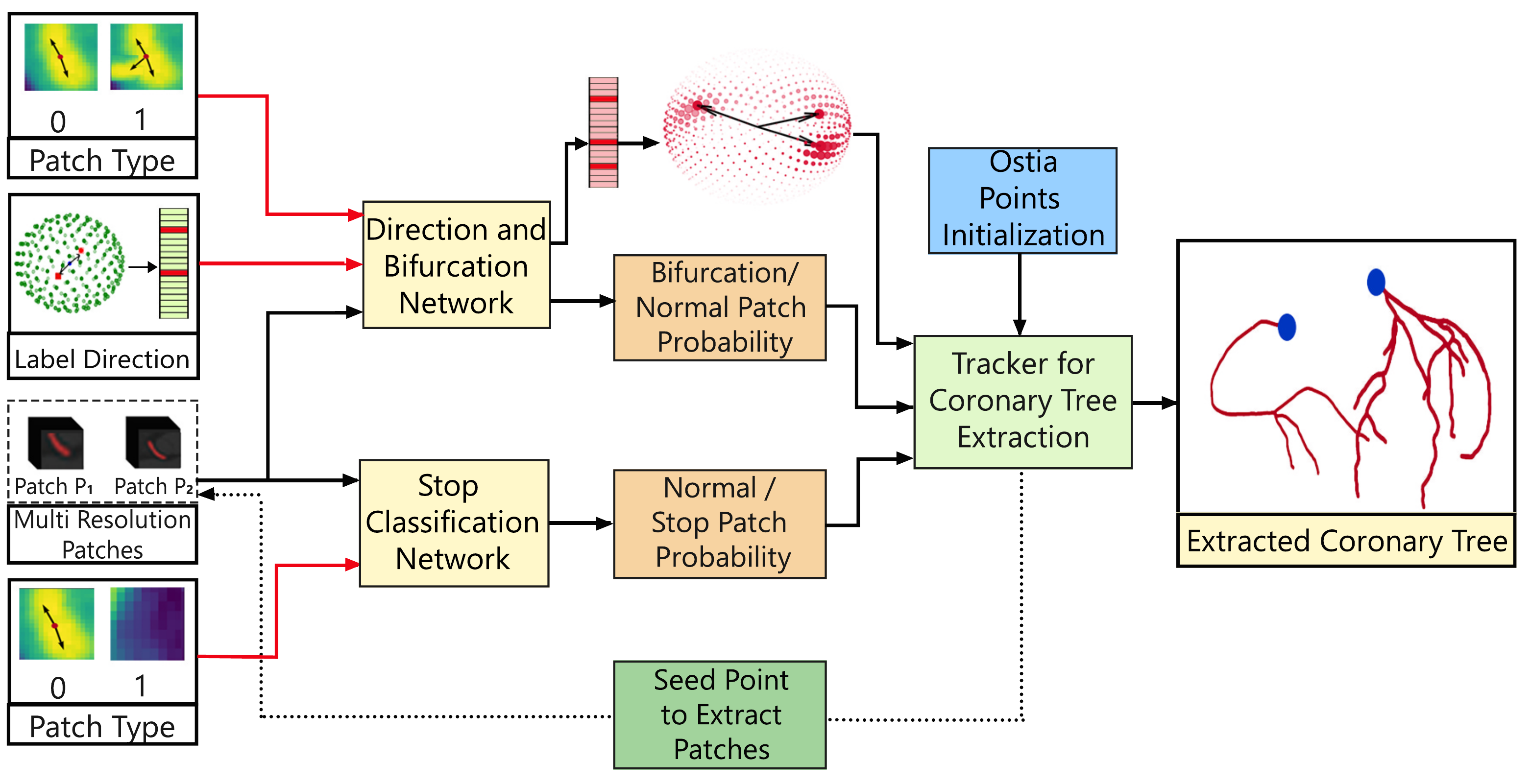}}
\caption{Overview of the proposed method. The red lines represent the inputs given to the direction and bifurcation classification model (DBC-Net) and the stop patch classification model (STC-Net) during the training phase along with the multiresolution patches $P_1$ and $P_2$. }
\label{figure : overview}
\end{figure*}

We propose a novel 3D CNN-based algorithm that is able to extract the entire coronary tree automatically. This approach does not require any pre-processing step or handcrafted filters. No user interaction is required to obtain the entire coronary tree. In contrast to the CNN approach by \cite{WOLTERINK201946}, bifurcations are also detected by the CNN and consequently the resulting directions are predicted by the CNN which make it possible for the entire coronary tree to be extracted by a seed point placed automatically anywhere on the coronary tree instead of requiring one seed point per vessel. The termination of the tracking in our proposed method is guided by another 3D CNN which prevents premature termination. Figure \ref{figure : overview} shows overview of the entire pipeline of the proposed algorithm.

\section{Method}

The training dataset for the direction classification and bifurcation detection model (\textbf{DBC-Net}) and stop patch classification model (\textbf{STC-Net}) consists of 3D isotropic patches \textbf{$P_1$} of resolution $0.5 \times 0.5 \times 0.5 \thinspace \text{mm}^3$ and \textbf{$P_2$} of resolution $1 \times 1 \times 1 \thinspace \text{mm}^3$. These training patches are centered at a location \textbf{x} in a CCTA image in the vicinity of an annotated centerline point. The radius of the annotated coronary arteries ranges from 0.179 mm to nearly 3.55 mm near the ostium in the in-house dataset. As a rule of thumb, the size of the patches should be small enough to not lose the local context in the smaller arteries but it should be also be big enough to be able to determine the direction in the broader portion of the arteries. This implies  the approximate minimum patch size to be $\frac{\textbf{3.55}  \thinspace \times  \thinspace \textbf{2} }{\textbf{0.5}}$ = 14. In order to cover artery sections with a large diameter, we choose the patch size of 19. It is small enough to allow fast forward and backward pass as well as sufficiently large to encapsulate the whole context of the coronary artery information. 

\begin{figure}[h!]
\centerline{\includegraphics[width=0.4\columnwidth]{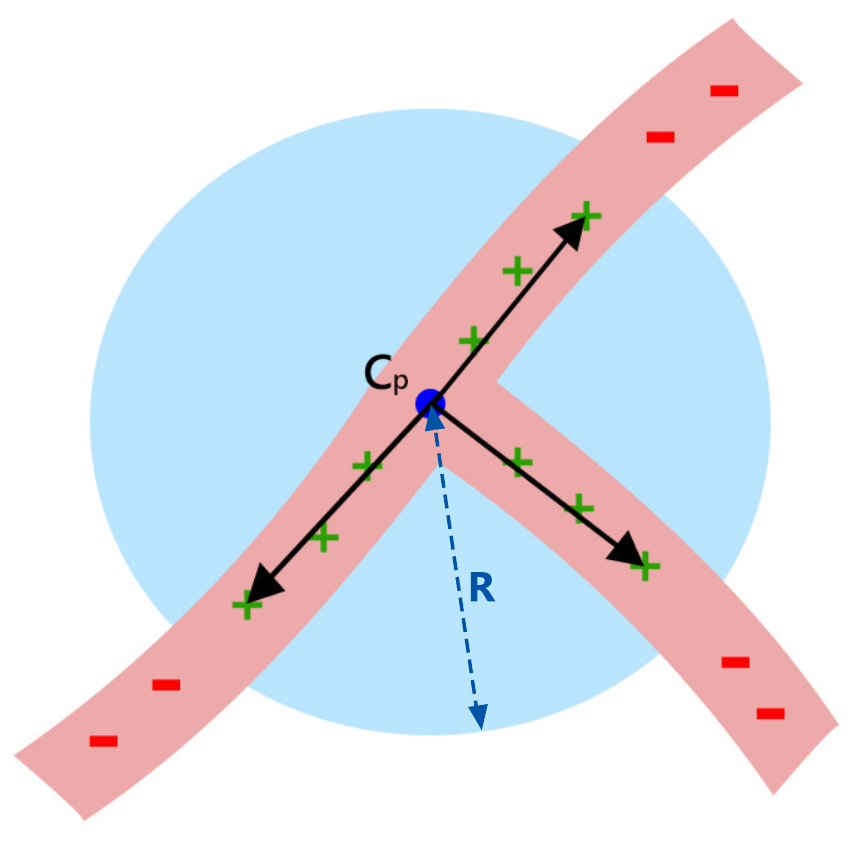}}
\caption{2D projection of the 3D sphere of radius $R$ placed at center of the patch $c_p$ for getting label direction vectors. The annotated centerline points inside the sphere are indicated by a + sign and the ones outside are indicated by a - sign. The label direction vectors are obtained by detecting the sign changes.}
\label{figure:label assignment}
\end{figure}

\begin{figure}[t!]
\centerline{\includegraphics[width=0.4\columnwidth]{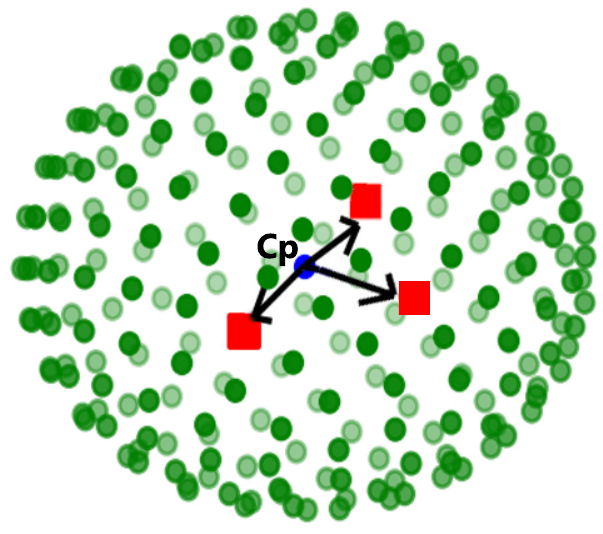}}
\caption{The green dots on the sphere $\boldsymbol{S_d}$ correspond to the $\boldsymbol{N_d}$ admissible movement directions. The center of the patch $\boldsymbol{c_p}$ is denoted by a blue dot and the red squares indicate the closest points on the sphere grid associated with the direction vectors which are assigned the value 1. The remaining grid points are assigned the value 0.}
\label{figure:sphere_label_assignment}
\end{figure}

The direction vectors from the center of the patch $c_p$ to the adjacent centerline points need to be determined to guide the tracking algorithm. The label direction vectors are obtained by placing a sphere of radius $R$ at the center of the patch as shown by Figure \ref{figure:label assignment}. The annotated centerline points within the sphere are designated as positive and those outside the sphere as negative. We determine the exit points of the arteries contained in the sphere by observing the sign changes associated with each artery. If there is a bifurcation, there will be three exits (sign changes) from the sphere and there will be only two exits in the normal case. The sphere radius should be large enough in order to detect three sign changes associated with the occurrence of the bifurcation. However, if  $R$ is made too large, the bifurcations will be detected well before the patch center $c_p$ is at the bifurcation point. This would allow tracked centerlines to branch prematurely before only to be joined later. The radius $R$ was fixed to 1.5 mm. 
The direction vectors obtained are then discretized on a unit sphere $S_d$ placed at the center of the patch. Approximate equidistant discrete grid on this sphere is obtained using Spherical Fibonacci Mapping \cite{Keinert2015}. Each grid point corresponds to an admissible movement direction. Given a set of direction vectors related to a specific center point, the point on the discrete sphere grid which makes the minimum angle with the corresponding direction vector is assigned the value 1. All other points on the sphere grid which do not have any direction vector associated with them are assigned the value 0. Figure \ref{figure:sphere_label_assignment} shows how the label direction vectors are associated with discrete locations on the sphere. Finally, the vector encoding the movement directions is normalized to unit length. The problem of determining the direction vectors is then simply reduced to the classification of discrete locations on the unit sphere. The number of discrete locations $N_d$ on the unit sphere $S_d$ is fixed to 1000.

\begin{figure}[h!]
\centerline{\includegraphics[width =0.8\columnwidth]{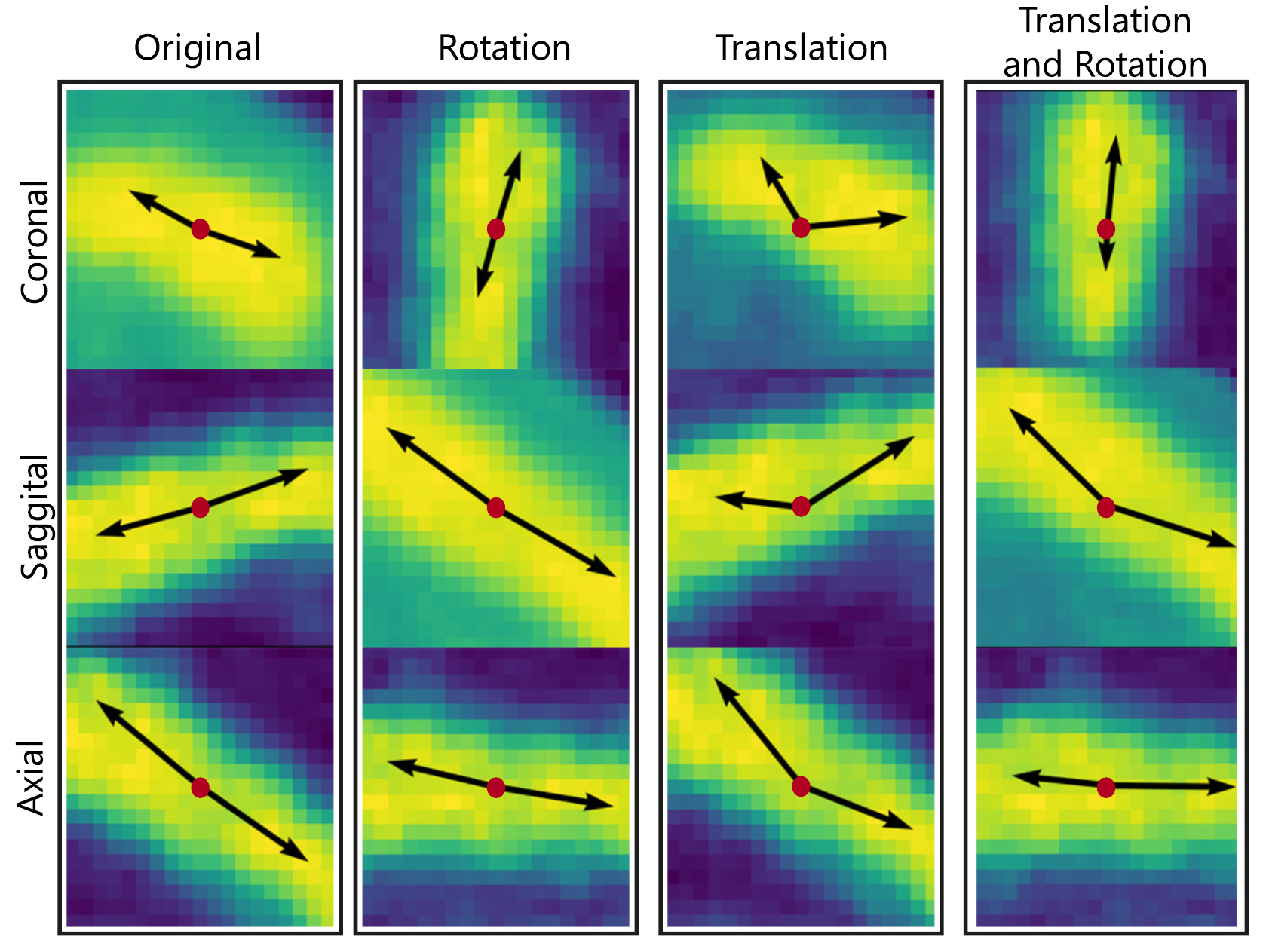}}
\caption{Maximum intensity projections (MIP) in coronal, axial and sagittal views of the 3D patch in order to visualize augmentation. The direction vectors are shown after applying 3D rotation and 3D translation augmentation.}
\label{figure : augmentation}
\end{figure}

\subsection{Augmentation}
We use various augmentation strategies during training in order to improve the overall robustness of our tracker.
Firstly, augmentation by randomly generated translations was introduced to teach the tracker how to recover from centerline deviations. Translation augmentation is introduced by adding a small deviation  $\Delta_t$ to the center of the patch $c_p$ and extracting the patch at this new translated center $C_t$ = $c_p$ + $\Delta_t$. This deviation should not be so large that the artery is no longer within the field of view. It should be ensured that at-least half of the artery is visible in the patch. Hence, the translation augmentation is applied with respect the radius of the artery at the annotated centerline point. The amount of applied deviation is $\Delta_t$ = $\lambda_t \thinspace \times \thinspace radius$ where $\lambda_t$ is uniformly sampled from the interval [0,1].
A small translation may result in a drastic change in direction vectors. The direction vectors are highly sensitive to the center of the patch $c_p$. Consequently, the label direction vectors are determined with respect to a pseudo center $C_{pseudo}$ which is closer to the original center in order to dampen the effect of the translation on the direction. 

\begin{equation}
C_{pseudo}  = 0.8 \cdot c_p + 0.2 \cdot \Delta_t
\end{equation}

Patches $P_1$ and $P_2$ are given as input to the DBC-Net  as shown in Figure \ref{figure: architecture}. These patches are extracted at the translated center $C_t$ and the label direction vectors are determined with respect to the pseudo center $C_{pseudo}$.
Rotational augmentation is also introduced by rotating the 3D patches around the center randomly around the three axis $(\phi_x, \phi_y, \phi_z)$. The label direction vectors are obtained before applying the rotational augmentation. The rotation of the patches is incorporated in the corresponding labels by applying the same rotation to the label direction vectors. The rotated direction vectors are then associated with discrete directions on the sphere $S_d$. Figure \ref{figure : augmentation} shows how the label direction vectors are transformed after applying augmentation.

\subsection{Ostium Points for Algorithm Initialization}
In order to initialize the fully automatic centerline extraction of the coronary arteries, two seed points corresponding to the left and right coronary trees need to be obtained automatically. The algorithm can be initialized by selecting a point which may be located anywhere on the coronary artery tree. An algorithm based on Model Based Segmentation (MBS) was used to determine the left and right ostium origin points from the aorta. These points were used for initialization of the tracking algorithm. The spatial location of the ostia landmarks is derived from the mesh topology \cite{Ecabert2008}.

\begin{figure}[h!]
\centerline{\includegraphics[width =0.55\columnwidth]{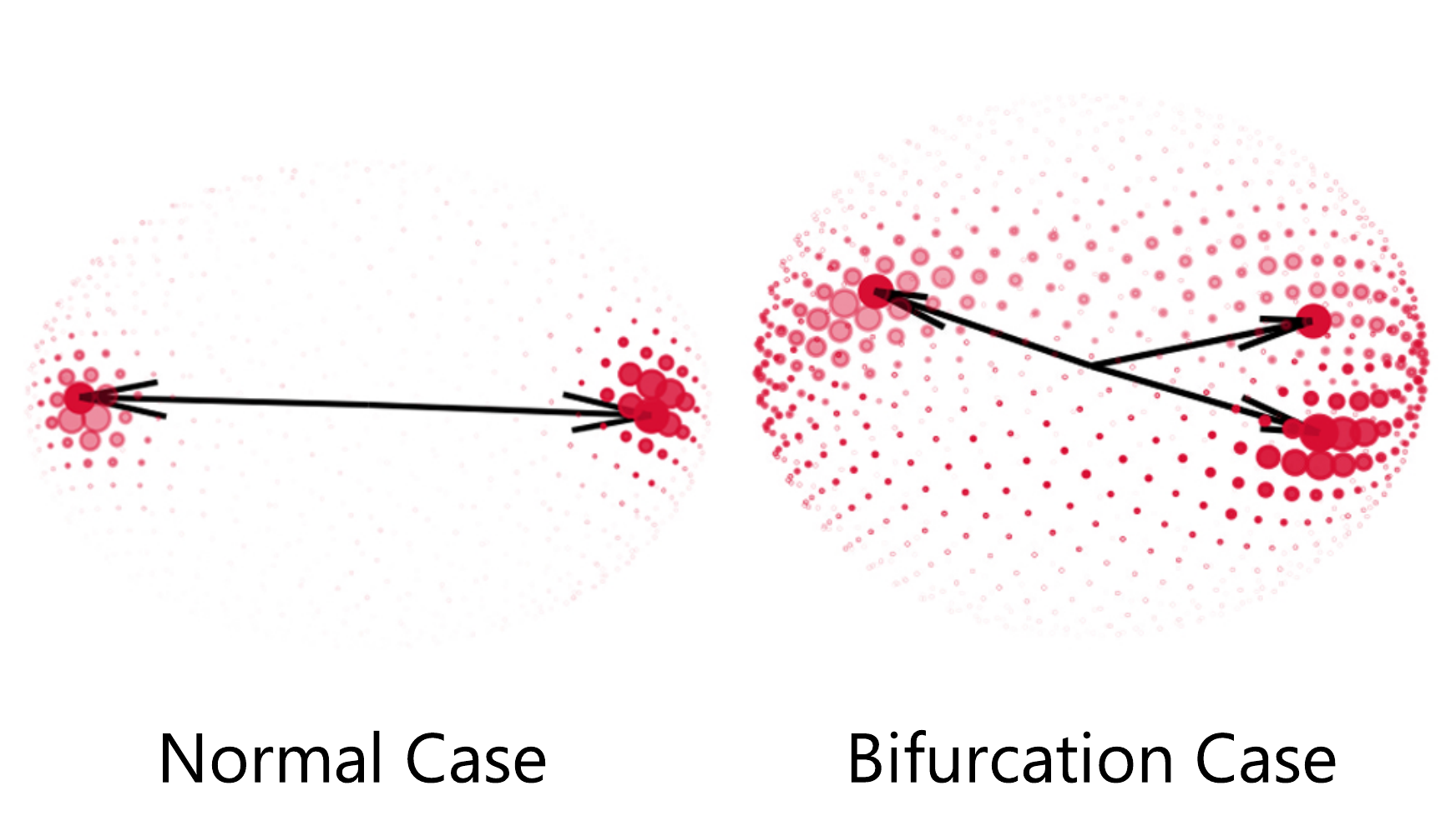}}
\caption{The output response obtained from the direction layer $L_D$ of DBC-Net. Two or three peaks are observed in the response according to the patch type. The resulting direction vectors obtained are also shown in black.}
\label{figure : output}
\end{figure}

\subsection{Bifurcation Prediction}
The entire left and right coronary tree can be traced by using one seed point each placed anywhere if all of the bifurcations are correctly detected by an algorithm. The accurate classification of patch type as bifurcation or normal is essential to the tracking of entire coronary tree based on a single seed point. Depending on this prediction, the number of direction vectors obtained will be two or three respectively. Hence, the subsequent network will also predict the bifurcation type in order to facilitate the tracking procedure.

In our training set, uniformly sampling center points from the coronary trees resulted in a rare occurrence of patches containing bifurcations. We utilized the strategy of  \textbf{Importance Sampling} in order to assure that 20\% of the patches in a mini-batch include bifurcation.

\begin{figure*}[h!]
\centerline{\includegraphics[width=\textwidth]{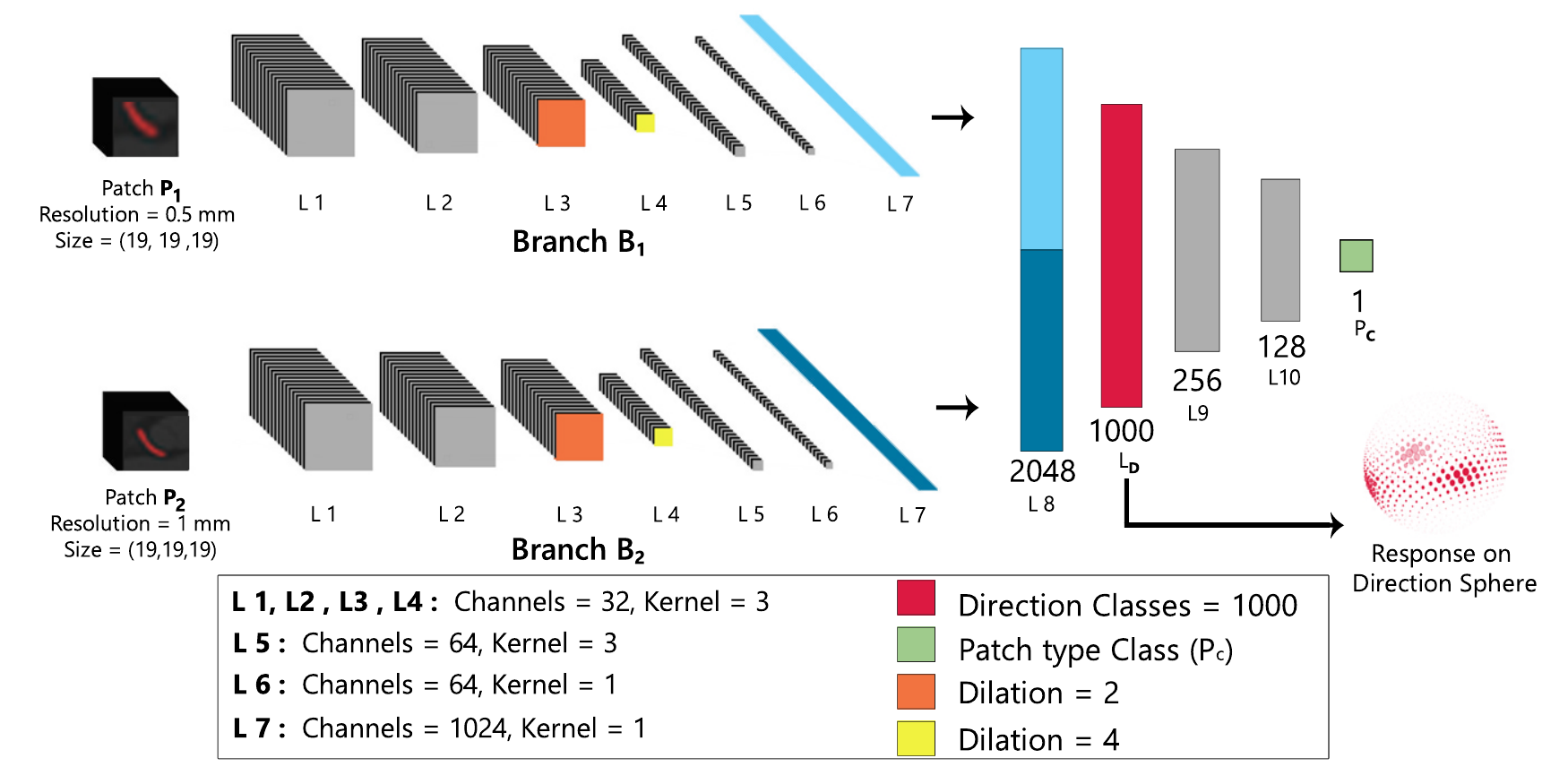}}
\caption{The proposed dual pathway multiresolution architecture proposed for simultaneous direction and patch type classification. Patches $P_1$ and $P_2$ of size $19 \times 19 \times 19$ with resolution $0.5 \times 0.5 \times 0.5 \thinspace \text{mm}^3$ and $1 \times 1 \times 1 \thinspace \text{mm}^3$ respectively are fed to the network to get direction class predictions on the direction sphere $S_d$. The direction layer $L_D$ is then followed by a series of linear layers to get a single patch type prediction $P_C$ (normal or bifurcation).}
\label{figure: architecture}
\end{figure*}

\subsection{Direction and Bifurcation Classification Network (DBC-Net)}
We propose a combined approach for classifying directions to the neighboring centerline points from patch center $c_p$ on a unit sphere $S_d$ having $N_d$ discrete directions, as well as patch type classification $P_{c}$ (Normal or Bifurcation). The employed CNN network consists of a Deep Medic-based architecture \cite{Kamnitsas2016}. Figure \ref{figure: architecture} shows that the proposed architecture has two branches $B_1$ and $B_2$ that take 3D patches $P_1$ and $P_2$ with resolutions $0.5 \times 0.5 \times 0.5 \thinspace \text{mm}^3$ and $1 \times 1  \times 1 \thinspace \text{mm}^3$ as input respectively. Each branch consists of 7 3D convolutional layers with kernel size of 3. Layers $L_3$ and $L_4$ use dilated convolutions with the spacing of 2 and 4 between the kernel points respectively. After each convolutional layer, 3D batch normalization was employed. At the end of these 7 layers, the output of the branches $B_1$ and $B_2$ is concatenated in Layer $L_8$. The output of $L_8$ is then reduced to the number of direction classes $N_D$ to form the direction class layer $L_D$. The Layer $L_D$ is then subjected to two linear layers $L_9$ and $L_{10}$ to finally get the patch type classification $P_C$.

ReLU activation functions are used after all layers except the layers $L_D$ and $P_c$. The patch type class layer $P_c$ uses a sigmoid activation function and the direction class layer $L_D$ uses a softmax activation function. \textbf{Binary cross entropy loss $(BCE_{patch})$} is used for the patch type classification and \textbf{categorical cross entropy loss $(CE_{direction})$} is used for the direction classification. The combined loss function used to train the network is as follows:

\begin{equation}
    Total \thinspace Loss = CE_{direction} + \lambda_b \cdot BCE_{patch}
\end{equation}

The weighting factor $\lambda_b$ is fixed at 5. The other hyper-parameters tuned for this set up include learning rate of 0.0001 with Adam optimizer and mini-batch size of 64.

\begin{figure}[th!]
\centerline{\includegraphics[width =0.5\columnwidth]{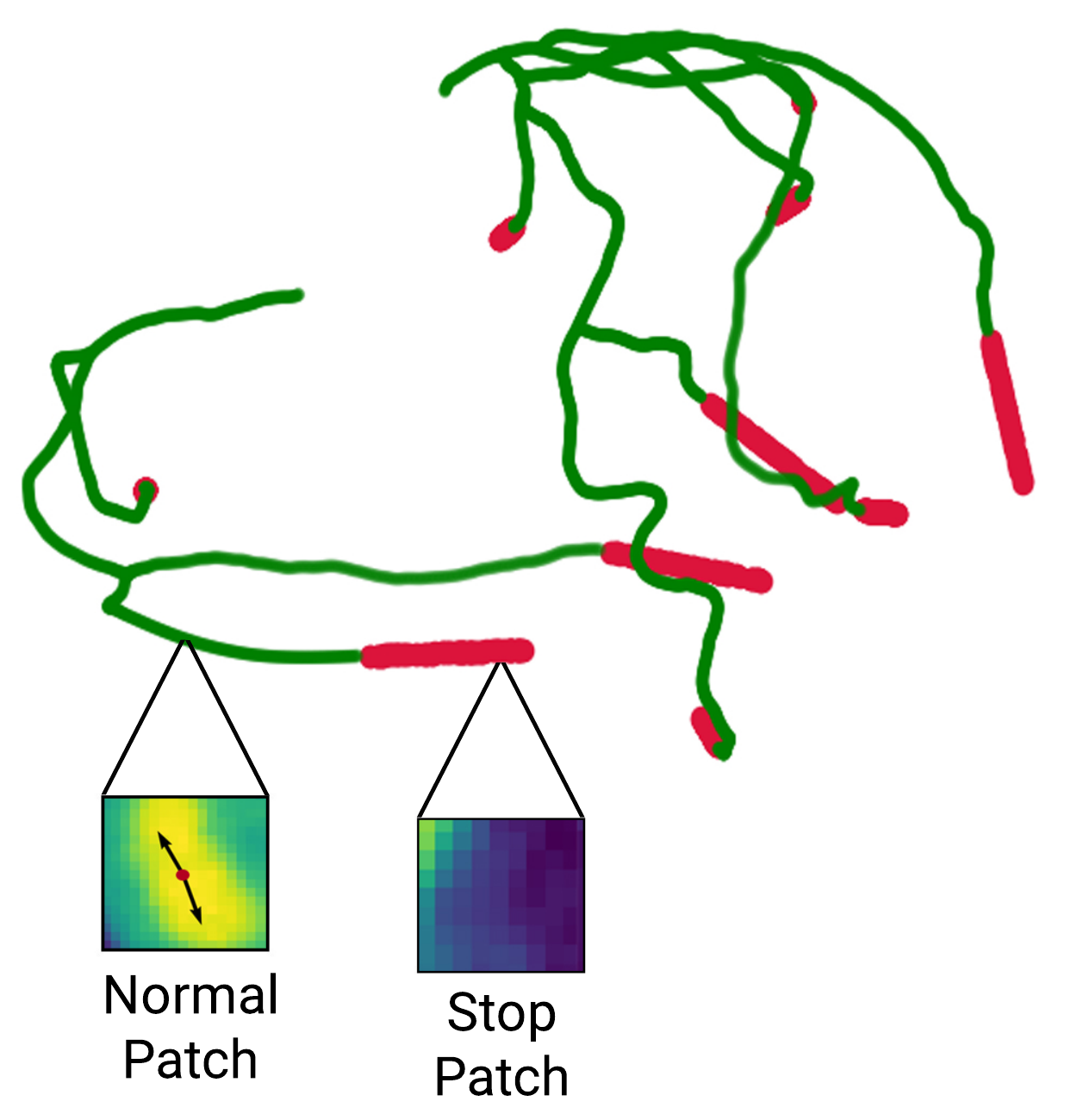}}
\caption{The patch labels used for training the stop patch classification model. The ``Normal" class patches correspond to the green portion in the figure. The ``Stop" class patches correspond to red portion beyond the last annotated centerline point.}
\label{figure: stop_patch}
\end{figure}

\lstinputlisting[label = {listing : lst1},float=*, basicstyle=\footnotesize, caption={Pseudo code of the tracking algorithm. $get\_direction\_vectors$ function returns the direction vectors to the neighboring centerline points taking the direction response and bifurcation prediction from the DBC-Net as input. The \textit{centerline} list contains the tracked coronary tree at the termination of this algorithm.},frame=single]{AuCoTrack.pseudo_code}
\subsection{Stop Patch Classification Network (STC-Net)}
In order to terminate the tracking algorithm once the end point of a coronary artery has been reached, we propose to train a separate 3D CNN. The same architecture as shown in Figure \ref{figure: architecture} is used to train the stop patch classification model (\textbf{STC-Net}). The endpoints of coronary arteries can be quite ambiguous. In order to get patches corresponding to the stopping criteria, we sample points beyond the end of the coronary arteries. This is achieved by using the direction vector obtained by subtracting the penultimate centerline endpoint from centerline endpoint. Points beyond the endpoint up to 5 mm are sampled and labelled as stop patch type. All the other centerline points are labeled as normal patch type. Only binary cross entropy loss for stop patch type classification is employed for training the network. The overall stopping criteria is based on the predictions by the STC-Net and the entropy of the direction prediction response by the DBC-Net. \cite{WOLTERINK201946}'s stopping criterion is solely based on moving average entropy which results in premature termination as well as leakage in some of the cases. Our combined stopping heuristic tries to  solve the issue of premature termination in the presence of plaque and stenosis.

\subsection{Tracking Implementation}
The tracking starts by obtaining the seed points, one for each coronary tree, from the ostium initialization module. These seed points are added to an active queue. We continue the tracking until there are no points in the active queue. We obtain two patches $P_1$ and $P_2$ of resolution $0.5 \times 0.5 \times 0.5 \thinspace \text{mm}^3$ and $1 \times 1 \times 1 \thinspace \text{mm}^3$ respectively centered at the point popped from the queue. These patches are fed to the DBC-Net and STC-Net. The STC-Net outputs the probability of the patch being a stop patch or normal patch. 

The DBC-Net determines the direction predictions on the unit sphere $S_d$ as well as classifies if the given patch contains a bifurcation or not. The DBC-Net learns to predict some relatively high values near the correct direction class as, for example, shown by Figure \ref{figure : output}. We observed that high probabilities were assigned to  the neighbors of the correct direction class as well. Smoothing with a Gaussian kernel of size 16 was applied in order to get rid of the noise. Once the predictions are smoothed out, we detect two or three peaks depending on the patch type classification.

Depending on the direction response prediction, the direction $D_1$ is obtained by taking into account the previously tracked centerline point. If this is the first point being extracted, we take the maximum of the direction response as $D_1$. If a centerline point has been previously extracted, we take into account the previous direction $D_{prev}$ used to obtain this patch. The angle between $D_1$ and $D_{prev}$ should be less than 60$^{\circ}$ in order to make sure that the tracker always proceeds forward. The maximum response obtained in this constrained field of view is labeled as $D_1$. The second direction $D_2$ should be at least 110$^{\circ}$ farther from $D_1$ in order to make sure that the opposite direction is correctly tracked. The third direction $D_3$ should be at least 40$^{\circ}$ farther from the first and second responses. In case the patch type is normal, only $D_1$ and $D_2$ are determined. The candidate points $S_{cand_i}$ are obtained from the direction vectors $D_i$ as follows:

\begin{equation}
    S_{cand_i}  =  S_{point}\thinspace + \thinspace \Delta_{step} \cdot D_i
\end{equation}

The parameter \textbf{i} varies from 1 to 2 in normal case and from 1 to 3 in case a bifurcation has been detected. Moreover, $S_{point}$ represents the current patch center and ${\Delta_{step}}$ is the step size. The distance between all the candidate points and already finalized centerline points is determined. Candidate points with a distance $> 0.5 ~\Delta_{step}$ are added to the active queue.

A combined criterion based on the predictions of the STC-Net and entropy determined from the direction prediction response of the DBC-Net is used to terminate the tracking. If the entropy exceeds a threshold of 0.8 or the stop patch probability goes above 0.5, the stop counter is incremented by one. If the none of these two conditions are satisfied, the counter is reset to zero. If the stop counter exceeds 3, the active point is not added to the list of tracked centerline points and tracking terminates. 

\begin{figure}[bh!]
\centerline{\includegraphics[width=0.4\columnwidth]{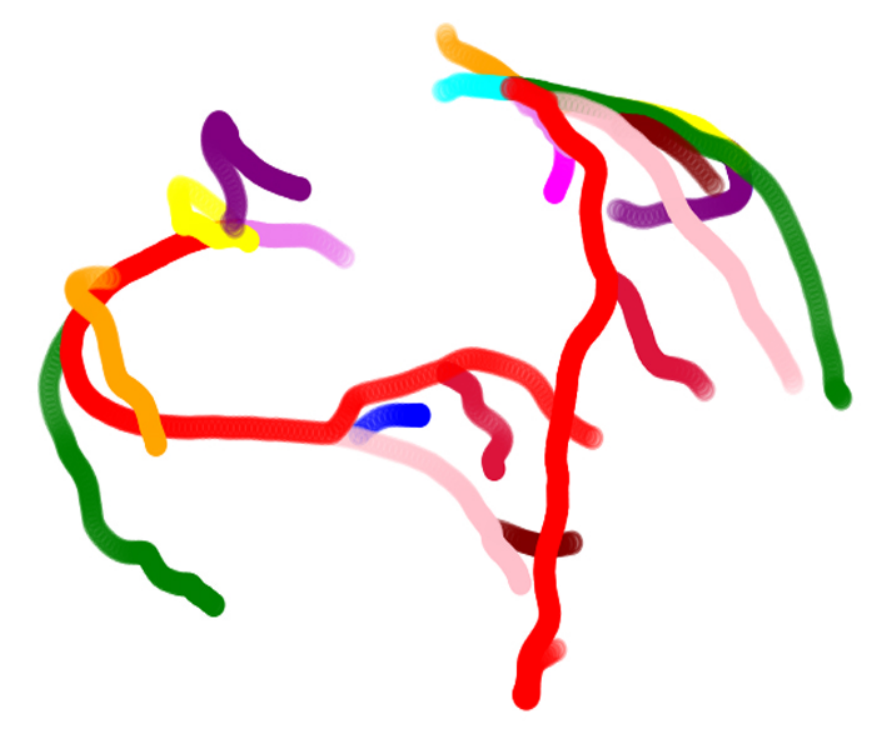}}
\caption{Different vessels in the coronary tree obtained from the tracked result. Each color represents a different coronary artery.}
\label{figure : topological}
\end{figure}

Information related to the previous direction and stop counter is kept in the active queue along with the 3D point coordinates. It is also important to keep track of the separate segments and their parents in the queue. The segment terminates at each bifurcation point. Listing \ref{listing : lst1} shows the simplified pseudo code for the tracker implementation. Figure \ref{figure : topological} shows how individual vessels can be obtained making use of the segment information stored during tracking.

\begin{figure}[th!]
\centerline{\includegraphics[width=0.9\columnwidth]{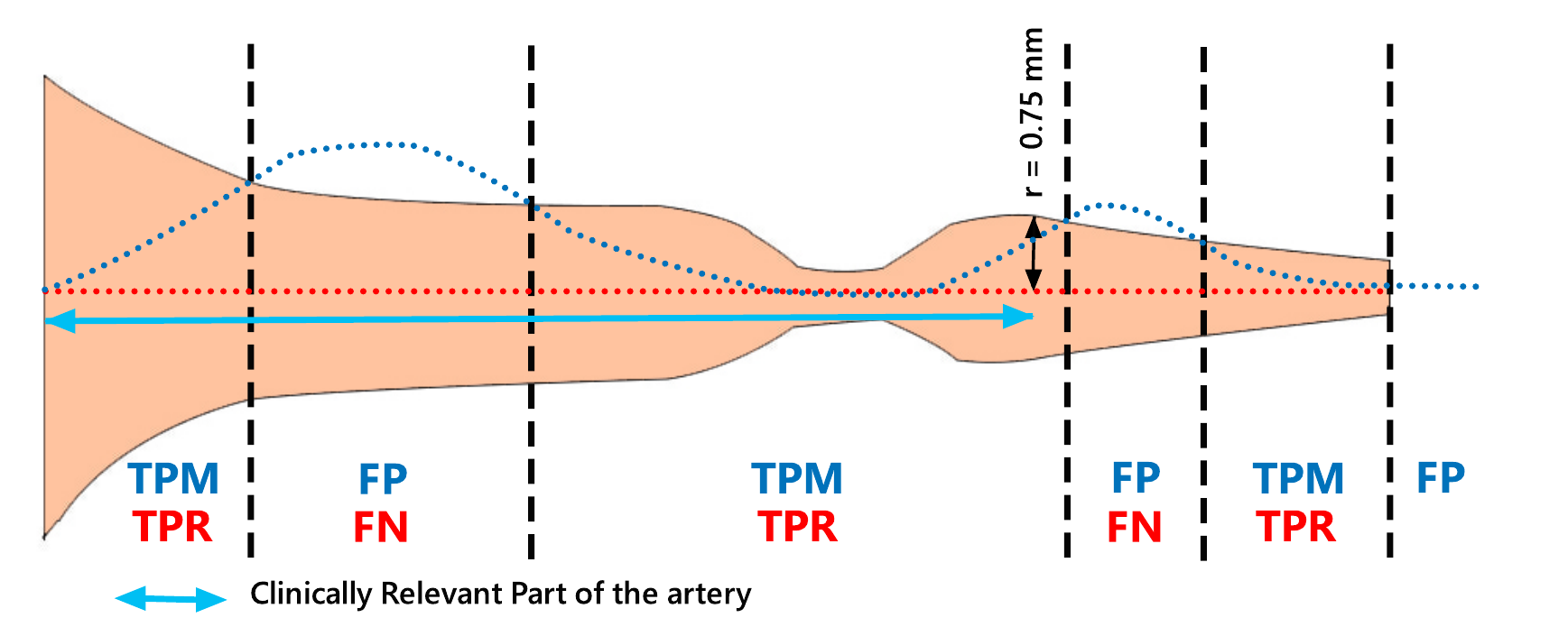}}
\caption{True Positive Reference (TPR), True Positive Measured (TPM), False Positive (FP) and False Negative (FN) associated with different parts of the tracked (blue dotted line) and reference centerline (red dotted line). Clinically relevant part of the vessel is also shown in the figure. More details about metric calculation can be found in \cite{Schaap2009701}.}
\label{figure : metrics}
\end{figure}

\subsection{Evaluation Measures}
A prerequisite to the evaluation of all the metrics is the conformity in the spacing between the tracked centerline points and the ground truth centerline points. The ground truth annotations of the coronary arteries and the tracked arteries are resampled uniformly to obtain same spacing between consecutive points \cite{Schaap2009701}.
A point on the ground truth centerline is labeled as True Positive Reference \textbf{(TPR)} if a tracked centerline point is present within the corresponding annotated radius and it is labeled as False Negative \textbf{(FN)} otherwise.
A point on the tracked centerline is labeled as True Positive Measured \textbf{(TPM)} if a ground truth centerline point is present within the corresponding annotated radius and it is labeled as False Positive \textbf{(FP)} otherwise. Figure \ref{figure : metrics} shows how TPR, TPM, FP and FN are obtained in terms of tracked and reference centerlines. 
The points towards the distal end of the coronary arteries may be ambiguous and not clinically relevant. The endpoint of the clinically relevant part of each artery is defined as the most distal point of the vessel with an radius greater than 0.75 mm.

\textbf{Sensitivity} determines how much of the ground truth coronary tree has been correctly tracked by the algorithm. A sensitivity value of 1 indicates that the entire coronary tree has been covered by the centerline extraction algorithm.

\begin{equation}
\text{Sensitivity}  = \frac{TPR}{TPR + FN}
\end{equation}

The number of annotated coronary arteries varies from 4 to 20 in the CCTA images in the in-house dataset. In an effort to obtain the entire coronary tree, the algorithm will also track the arteries that have not been annotated. However, a check should be maintained to see that the algorithm doesn't detect many spurious vessels. Hence, the deviation from the coronary reference tree is kept in check in terms of \textbf{False Positive Rate} (FPR). For calculating the Sensitivity and False Positive Rate, we set the threshold radius to 1 mm.

\begin{equation}
\text{FPR} = \frac{FP}{TPM + FP}
\end{equation}

Overlap measure as defined in equation \ref{overlap} similar to dice in segmentation. \textbf{Average Overlap} (OV) takes the entire reference and extracted coronary artery into consideration. \textbf{Clinically Relevant Overlap} (OT) calculates the overlap only for the clinically relevant part of the artery. \textbf{Overlap until First Error} (OF) calculates the portion of the overlap accurately tracked until the first error occurs \cite{Schaap2009701}.

\begin{equation}
\text{Overlap} = \frac{TPR + TPM}{TPR + TPM + FN + FP}
\label{overlap}
\end{equation}

The deviation of the extracted points from the reference centerline points is determined only for regions of the reference tree which are labelled as True Positive Reference. The average of the Euclidean distance between the reference centerline points and the nearest tracked centerline point determines the \textbf{Accuracy Inside} (AI).

\section{Results}
Fully automatic coronary centerline extraction methods extract the entire coronary tree without requiring any manually placed seed point to be provided for the vessels. The quantitative analysis of the extracted centerlines is performed individually for each coronary artery. The MICCAI CAT08 dataset provides for each case a point in the distal end of the coronary artery for selection. In case the coronary artery centerline is not present in the distal end, another point is  provided in the proximal part of the artery which can be utilized for coronary artery selection.\footnote{\url{       http://coronary.bigr.nl/centerlines/about.php/rules.php}} The evaluation guidelines of the challenge only allow the usage of one of these points. In order to keep the evaluation consistent, the quantitative analysis in in-house dataset is also performed by utilizing a point in the distal or proximal end of the coronary artery for selection.

\begin{figure*}[h!]
\centerline{\includegraphics[width=0.8\textwidth]{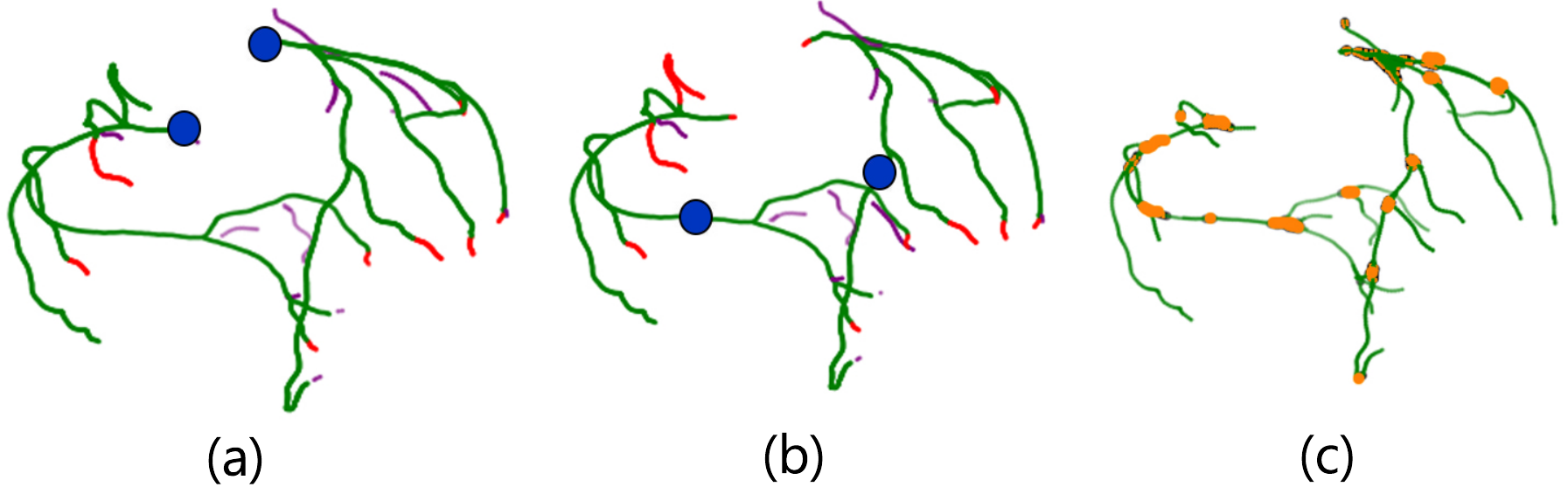}}
\caption{ Qualitative representation of the tracked result. (a) shows the coronary artery centerline extraction result when the algorithm is initialized by placing the seed points at the ostium. (b) shows the coronary artery centerline extraction result when the algorithm is initialized by placing the seed points in the middle of LAD and RCA. The green part of the extracted coronary tree indicates the portion tracked within the radius threshold of the annotated centerline points. The red part corresponds to the part missed by the tracking algorithm. The purple part corresponds to extra tracked arteries that are not present in the ground truth annotation. In (c), the bifurcation detection in orange is shown as overlay on the tracked coronary artery tree.}
\label{figure: qualitatie_plot}
\end{figure*}

\subsection{In-House Dataset}\label{sec:inhouse-data}
The in-house dataset consists of 43 CCTA images acquired from 9 different clinical which were annotated by clinical experts. The cases have been reconstructed around the 75\% cardiac phase and correspond to patients with suspected coronary artery disease. The dataset contains images from 64-slice Brilliance CT, 256-slice Brilliance iCT, Ingenuity CT, IQon Spectral CT scanners and a some images from SOMATOM Force CT scanners. The CCTA images in the dataset have a resolution ranging from  $0.25 \times 0.25 \times 0.33 \thinspace \text{mm}^3$ to  $0.48 \times 0.48 \times 0.80 \thinspace \text{mm}^3$ with a mean resolution of $0.40 \times 0.40 \times 0.43 \thinspace \text{mm}^3$. There is considerable variability in the  coronary arteries labeled for each case. The number of annotated coronary arteries per CCTA scan varies from 4 to 20. The mean number of annotated coronary arteries per CCTA scan in this dataset is 9. Depending on the number of annotated coronary arteries, the number of centerline points per case varies from 933 to 3200 with a mean of 1737. The in-house dataset in total contains 428 annotated coronary arteries. Four-fold cross validation has been performed in order to evaluate the proposed algorithm. The dataset was randomly shuffled and 33 CCTA images were used for training. The remaining 10 CCTA images were used for validation. We trained the DBC-Net for simultaneous direction classification and bifurcation detection as well as the model for the detection of stop patches. The seed point for the initialization of the tracker in order to obtain the centerlines for left or right coronary tree can be given anywhere on the coronary tree. However, the seed point for all the experiments was given near the ostium as this point can be obtained automatically from the model based segmentation.

\begin{figure*}[h!]
\centerline{\includegraphics[width=0.8\textwidth]{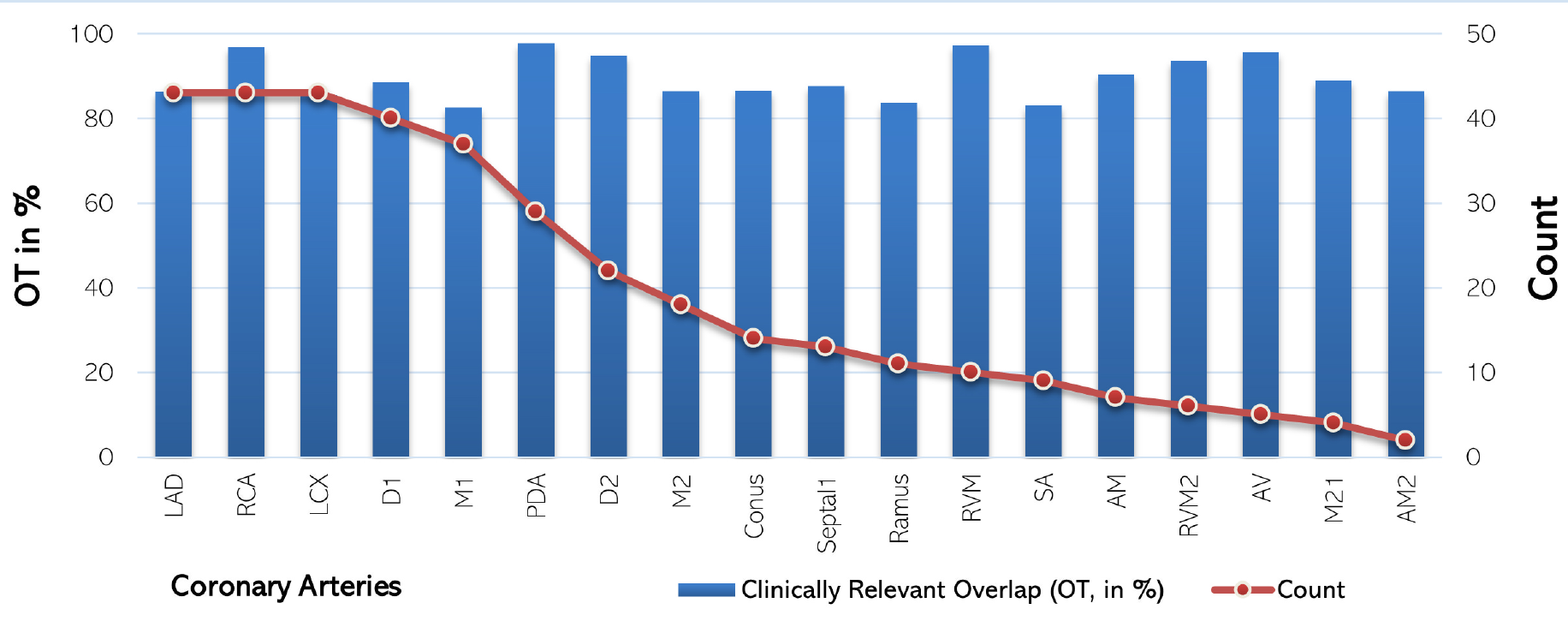}}
\caption{Clinically relevant overlap of all the arteries present the in-house dataset which occur more than 3 times. The number of occurrences of each artery is also shown in the plot. The horizontal axis shows the names of different coronary arteries, the left vertical axis shows their corresponding mean clinically relevant overlap obtained and the right vertical axis shows the number of times each artery is encountered in the in-house dataset.}
\label{figure: OT_arteries}
\end{figure*}

\begin{table}[t!]
\centering
\resizebox{\columnwidth}{!}{
\begin{tabular}{ccccc} 
\hline\hline
\textbf{No. of Resolutions} & \textbf{S}   & \textbf{OV}   & \textbf{OT}   & \textbf{AI}    \\ 
\hline\hline
\textbf{1}                  & 82.3          & 76.4          & 85.9          & 0.37           \\
\textbf{2}                  & \textbf{88.9} & \textbf{81.2} & \textbf{87.4} & \textbf{0.32}  \\
\textbf{3}                  & 87.4          & 78.6          & 86.5          & 0.34           \\
\bottomrule
\end{tabular}}
\label{table:num_resolutions}
\caption{The effect of varying number of resolutions levels in the DBC-Net in terms of total  sensitivity (S, in \%), overlap (OV, in \%), clinically relevant overlap (OT, in \%) and accuracy inside (AI, in mm). }
\end{table}

\begin{figure}[h!]
\centerline{\includegraphics[width=0.7\columnwidth]{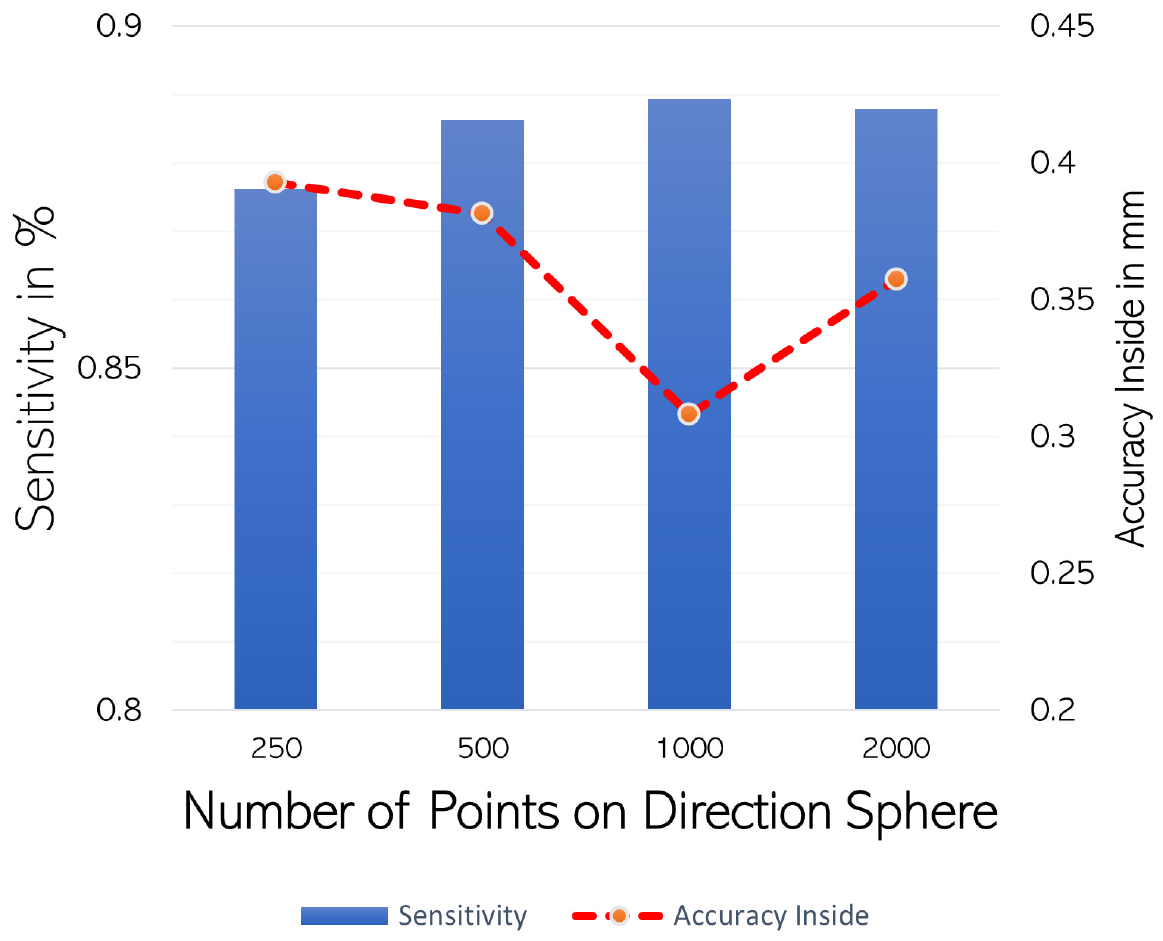}}
\caption{The effect of number of discrete direction points $N_d$ on the unit sphere  $S_d$ for direction classification in terms of sensitivity and accuracy inside.}
\label{figure: sphere_dir}
\end{figure}

\begin{figure}[h!]
\centerline{\includegraphics[width=0.7\columnwidth]{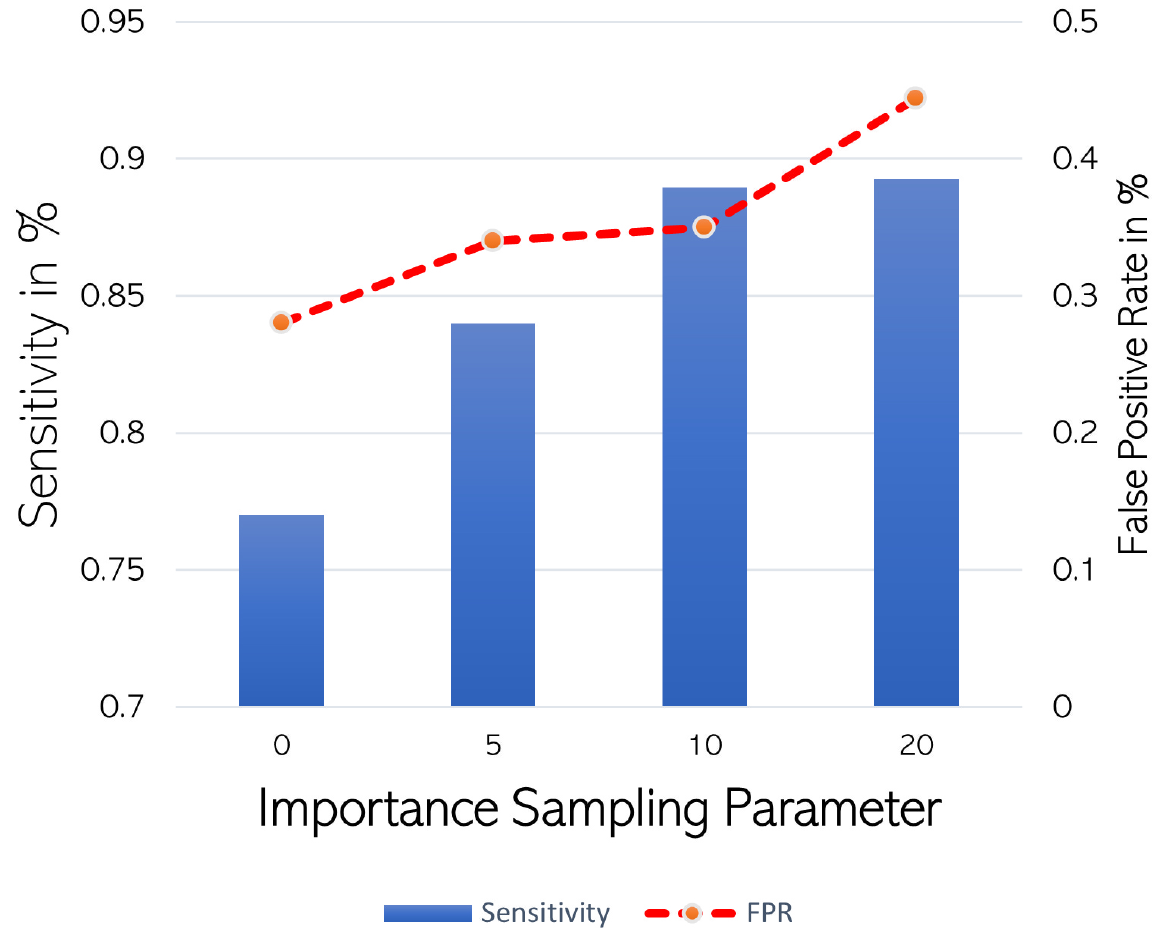}}
\caption{The effect of  importance sampling parameter while using a mini-batch of size 64 on sensitivity and false positive rate.}
\label{figure: imp_samp}
\end{figure}

Table 1 shows the overlap metrics and accuracy inside on the validation set when the number of input resolution levels and consequently the pathways in the architecture are varied. In case of a single pathway, only one resolution of $0.5 \times 0.5 \times 0.5 \thinspace \text{mm}^3$ is used.  In case of a dual pathway, two resolutions of $0.5 \times 0.5 \times 0.5 \thinspace \text{mm}^3$ and $1 \times 1 \times 1 \thinspace \text{mm}^3$ are utilized. In case of three pathways, three resolutions of $0.5 \times 0.5 \times 0.5 \thinspace \text{mm}^3$, $1 \times 1 \times 1 \thinspace \text{mm}^3$ and $1.5 \times 1.5 \times 1.5 \thinspace \text{mm}^3$ are employed. The dual pathway architecture with only two resolution levels performs better as indicated by overlap and accuracy metrics.
Figure \ref{figure: qualitatie_plot} (a) shows that the result of automatic coronary centerline extraction when the seed points for tracker initialization are placed at the left and right coronary ostium. Figure \ref{figure: qualitatie_plot} (b) shows that when the seed points are placed in the middle of LAD and RCA coronary arteries. The extracted coronary tree is almost similar in both the cases. The sensitivity obtained when the seed points are placed at the coronary ostia for fold 3 is 88.9\%  and it is 87.3\% when the seed points are placed in the middle of LAD and RCA. This shows that the seed point can essentially be placed anywhere on the coronary tree. Figure \ref{figure: qualitatie_plot} (c) shows the bifurcation detection overlayed on the tracking result. The orange color on the coronary tree indicates that a bifurcation has been detected at that centerline point by the DBC-Net which means that three direction vectors will be obtained to generate the candidate points.
Figure \ref{figure: sphere_dir} shows the effect of increasing the number of sphere direction points. As the number of the direction points increase, average accuracy inside and sensitivity drops after a maximum. Based on this observation, we fixed the number of direction points on the sphere at 1000 for further analysis

There is a data imbalance in the patches extracted from the CCTA images. Due to low number of patches with bifurcations, importance sampling is applied in order to ascertain that a percentage of the mini-batch during training contains bifurcation patches. Figure  \ref{figure: imp_samp} shows the effect of increasing the importance sampling factor while using a fixed mini-batch size of 64. As the importance sampling parameter increases, the sensitivity increases due to better detection of bifurcations. However, the detection of vessels not annotated in the dataset also increases. The importance sampling parameter is fixed at 10 in order to keep the false positive rate in check so that spurious vessels are not detected.

\begin{table}[h!]
\centering
\resizebox{\columnwidth}{!}{
\begin{tabular}{ccccc} 
\hline\hline
\textbf{Loss Function}                                                         & \textbf{S}   & \textbf{OV}   & \textbf{OT}   & \textbf{AI}    \\ 
\hline\hline
\begin{tabular}[c]{@{}c@{}}Softmax and\\Categorical Cross Entropy\end{tabular} & \textbf{88.9} & \textbf{81.2} & \textbf{87.4} & \textbf{0.32}  \\
\begin{tabular}[c]{@{}c@{}}Sigmoid and \\Binary Cross Entropy\end{tabular}     & 88.4          & 78.6          & 86.7          & 0.34           \\
\bottomrule
\end{tabular}}
\label{table:loss functions}
\caption{The comparison between two different types of loss functions  for direction classification in terms of sensitivity (S, in \%), overlap (OV, in \%), clinically relevant overlap (OT, in \%) and accuracy inside (AI, in mm).}
\end{table}

The final model uses softmax activation function for the direction classification layer (D) in the model and categorical cross entropy loss. Table 2 shows that this choice performs slightly better than sigmoid activation for the direction layer (D) and binary cross entropy loss. For the final hyper-parameter choice, we conducted a four-fold cross validation on randomly generated splits of the in-house dataset. The quantitative measures were averaged across the folds and we obtained an average sensitivity of 87.1\%,  clinically relevant overlap of 89.1\% and overlap of 80.4\% was obtained. An average accuracy inside of 0.34 mm was obtained which is within the average voxel size of $0.40 \times 0.40 \times 0.43 \thinspace \text{mm}^3$.

\subsection{CAT08 Dataset}
The MICCAI 2008 Coronary Artery Centerline Extraction Challenge (CAT08) dataset\footnote{\url{http://coronary.bigr.nl/centerlines/about.php/}} consists of 32 publicly available CCTA images comprising of 8 training  and 24 test CCTA images. The centerline annotations for test dataset are not available and the extracted centerlines can be evaluated only once on the evaluation framework.  CAT08 dataset contains images from 64-slice CT Siemens Scanner and dual source CT Siemens Scanner reconstructed to a resolution of $0.32 \times 0.32 \times 0.4 \thinspace \text{mm}^3$. Both the training and test set images were utilized as a test set for evaluating the performance of our algorithm on different scanners. Each image contains annotations for four arteries. The three fixed arteries present in all the CAT08 CCTA images include Left Anterior Descending Artery (LAD), Left Circumflex Artery (LCX) and Right Coronary Artery (RCA). However, the fourth artery in each case has arbitrarily been chosen. Since, fully automatic algorithms extract the entire coronary tree, there is a need to do a vessel by vessel evaluation \cite{Schaap2009701}. Hence, the CAT08 challenge provides with points in the distal end of the arteries that can be used to select the artery and evaluate metrics. If the entire coronary artery centerline extraction has not been successful, another point is also provided in the proximal end of the artery. Only one of these points may be utilized to select the artery.

\subsection{CAT08 Training Dataset}
The training dataset of CAT08 challenge was used as a test set in order to determine the performance of the proposed algorithm. The best model from the cross validation of the in-house dataset was used to extract centerlines for CAT08 training dataset. This model was not re-trained on CAT08 dataset. The tracker was initialized using ostium points derived from the Model Based Segmentation of the heart. The training dataset of CAT08 challenge consists of 8 CCTA images containing 32 annotated vessels. This dataset contains images of varying quality and calcium score.

Table \ref{table: table1} shows that an average overlap of \textbf{93.4\%}, clinically relevant overlap of \textbf{95.9\%} and overlap until first error of \textbf{76.5\%} was obtained for these 8 CCTA scans. All these CCTA scans have an image resolution of $0.32 \times 0.32 \times 0.4 \thinspace \text{mm}^3$.The average accuracy obtained was 0.36 mm which is approximately within the dimension of the one voxel. The average time taken to extract the entire coronary tree on a GTX 1080 GPU is 41 s. For all cases, 15 out of 16 vessels were automatically detected. One vessel from case 3 which was missed due to failure in corresponding bifurcation detection required an additional seed point in order to be detected. This is a good test of generalization of the algorithm as the model was trained on an in-house dataset of CCTA scans and evaluated on the CAT08 dataset.

\begin{table}[t!]
\centering
\resizebox{\columnwidth}{!}{
\begin{tabular}{cccccccl} 
\hline\hline
\textbf{No.}& \begin{tabular}[c]{@{}c@{}}\textbf{Image }\\\textbf{Quality}\end{tabular} & \begin{tabular}[c]{@{}c@{}}\textbf{Calcium}\\\textbf{~Score}\end{tabular} & \textbf{OV~}  & \textbf{OF~}  & \textbf{OT}   & \textbf{AI}   & \multicolumn{1}{c}{\textbf{T}}  \\ 
\hline\hline
0                                                                       & Moderate                                                                  & Moderate                                                                  & 94.2          & 77.7          & 95.1          & 0.4           & 55                              \\
1                                                                       & Moderate                                                                  & Moderate                                                                  & 97.3          & 99.4          & 99.6          & 0.32          & 39                              \\
2                                                                       & Good                                                                      & Low                                                                       & 98.3          & 99.7          & 100           & 0.31          & 43                              \\
3                                                                       & Poor                                                                      & Moderate                                                                  & 86.3          & 63            & 89.1          & 0.4           & 41                              \\
4                                                                       & Moderate                                                                  & Low                                                                       & 92.9          & 57.3          & 97.9          & 0.33~         & 31                              \\
5                                                                       & Poor                                                                      & Moderate                                                                  & 97.6          & 77.5          & 99.7          & 0.43~         & 33                              \\
6                                                                       & Good                                                                      & Low~                                                                      & 96.7          & 87.2          & 99.6          & 0.3~          & 36                              \\
7                                                                       & Good                                                                      & Severe                                                                    & 83.9          & 49.1          & 86.3          & 0.38          & 48                              \\ 
\bottomrule
\textbf{Avg}                                                        & \multicolumn{1}{l}{}                                                      & \multicolumn{1}{l}{}                                                      & \textbf{93.4} & \textbf{76.5} & \textbf{95.9} & \textbf{0.36} & \textbf{41}                              \\
\bottomrule
\end{tabular}}
\caption{Results of our method on CAT08 training set which was used as a test set. For each case, overlap (OV, in \%), overlap until first error (OF, in \%) and clinically relevant overlap (OT, in \%) ,average accuracy inside (AI, in mm), time taken for coronary tree extraction (T, in s)  along with subjective image quality and calcium score is shown.}
\label{table: table1}
\end{table}

\begin{figure}[b!]
\centerline{\includegraphics[width=0.7\columnwidth]{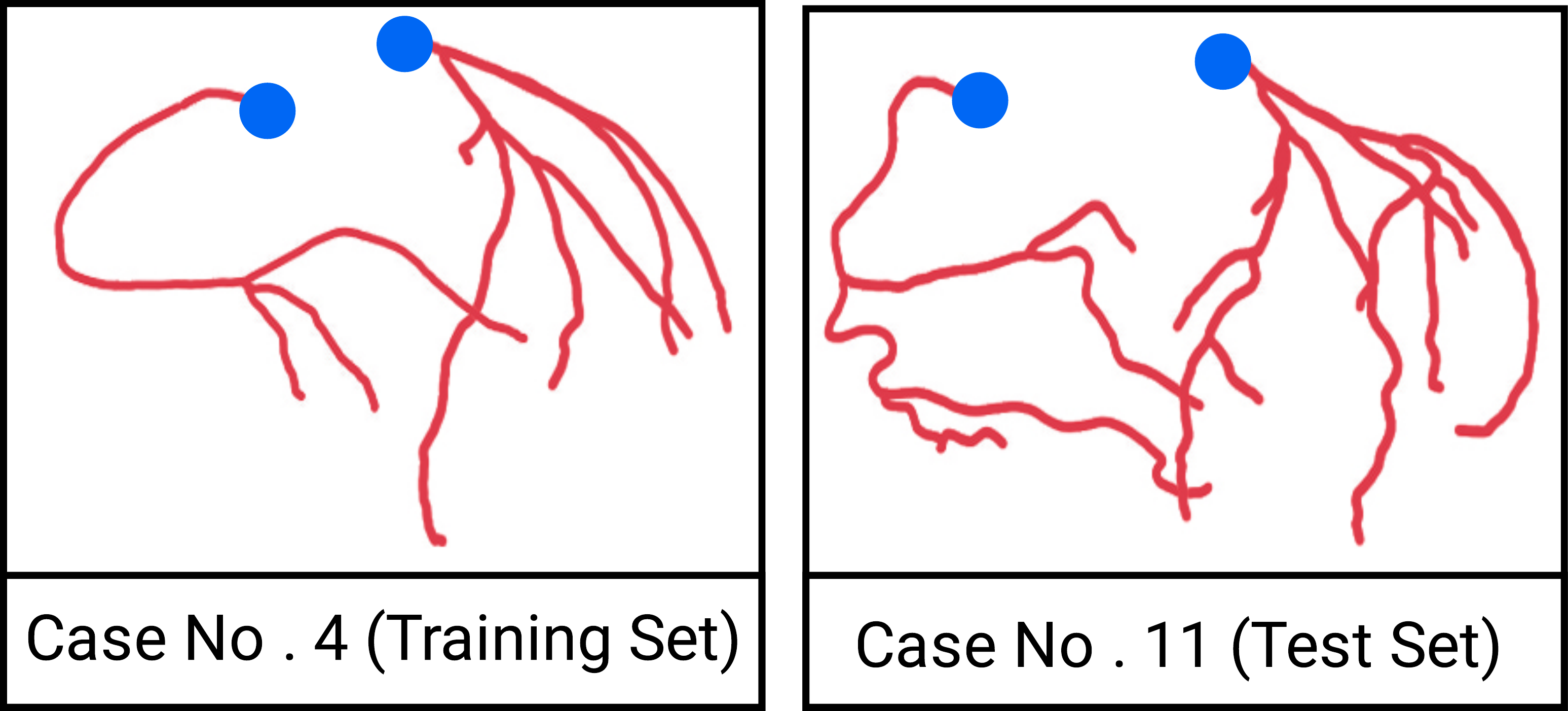}}
\caption{The left and right coronary trees extracted without any user intervention for two cases in training and testing set of CAT08 Challenge. The blue dots correspond to the ostia locations used for tracker initialization obtained automatically.}
\label{figure: result_CAT08}
\end{figure}

Figure \ref{figure: result_CAT08} shows the results of automatic coronary centerline extraction for case 4 from CAT08 training set and case 11 from CAT08 test set. The blue dots correspond to the ostia locations obtained from model based segmentation for tracker initialization. The proposed algorithm extracts the entire coronary tree while annotations for only 4 coronary arteries were provided in the ground truth.

\subsection{CAT08 Testing Dataset}
The CAT08 test dataset comprises of 24 CCTA scans of varying image quality and calcium scores. We tested our algorithm on these 24 CCTA images containing 96 vessels in order to benchmark the performance of our algorithm against methods available on the leaderboard of CAT08 challenge. Both the DBC-Net and the STC-Net were now trained on 43 cases of in-house dataset and 8 cases of CAT08 training dataset. These models were used to extract centerlines of the coronary arteries in the CAT08 test set which were then submitted to the evaluation framework online.

\begin{table}[t!]
\centering
\resizebox{\columnwidth}{!}{
\begin{tabular}{cccccccc} 
\hline\hline
\textbf{No.} & \begin{tabular}[c]{@{}c@{}}\textbf{Image }\\\textbf{Quality}\end{tabular} & \begin{tabular}[c]{@{}c@{}}\textbf{Calcium }\\\textbf{Scores}\end{tabular} & \textbf{OV}   & \textbf{OF}   & \textbf{OT}   & \textbf{AI}   & \textbf{T}     \\ 
\hline\hline
8            & Poor                                                                      & Low                                                                        & 84.6          & 48.7          & 91.1          & 0.46          & 31             \\
9            & Good                                                                      & Low                                                                        & 95.4          & 70.2          & 98.5          & 0.34          & 37             \\
10           & Moderate                                                                  & Moderate                                                                   & 94.8          & 91.4          & 97.3          & 0.36          & 36             \\
11           & Good                                                                      & Moderate                                                                   & 91            & 53.7          & 91.8          & 0.39          & 49             \\
12           & Good                                                                      & Moderate                                                                   & 89.1          & 29.4          & 94.2          & 0.38          & 55             \\
13           & Moderate                                                                  & Low                                                                        & 97.8          & 96.4          & 97.8          & 0.37          & 39             \\
14           & Moderate                                                                  & Severe                                                                     & 97.7          & 67.2          & 98.7          & 0.4           & 57             \\
15           & Moderate                                                                  & Moderate                                                                   & 96.1          & 100           & 100           & 0.34          & 65             \\
16           & Good                                                                      & Low                                                                        & 97.4          & 92.4          & 100           & 0.4           & 45             \\
17           & Poor                                                                      & Severe                                                                     & 91.6          & 68.3          & 95.8          & 0.41          & 33             \\
18           & Good                                                                      & Moderate                                                                   & 96.6          & 95.9          & 98.9          & 0.3           & 38             \\
19           & Moderate                                                                  & Moderate                                                                   & 95            & 100           & 100           & 0.34          & 35             \\
20           & Moderate                                                                  & Moderate                                                                   & 90.1          & 46.2          & 90.5          & 0.45          & 43             \\
21           & Good                                                                      & Low                                                                        & 96.7          & 97.2          & 99.3          & 0.36          & 36             \\
22           & Good                                                                      & Low                                                                        & 96.8          & 99.6          & 99.7          & 0.35          & 33             \\
23           & Moderate                                                                  & Moderate                                                                   & 98.3          & 97.4          & 99.2          & 0.36          & 47             \\
24           & Moderate                                                                  & Severe                                                                     & 93.3          & 52.2          & 95.3          & 0.3           & 35             \\
25           & Good                                                                      & Moderate                                                                   & 95.1          & 69.5          & 98.2          & 0.39          & 51             \\
26           & Poor                                                                      & Low                                                                        & 76.7          & 34.8          & 86.6          & 0.5           & 47             \\
27           & Good                                                                      & Moderate                                                                   & 85            & 55.3          & 85.6          & 0.42          & 47             \\
28           & Good                                                                      & Low                                                                        & 95.9          & 94.4          & 97.4          & 0.32          & 38             \\
29           & Poor                                                                      & Moderate                                                                   & 98.4          & 95.1          & 99.7          & 0.32          & 44             \\
30           & Good                                                                      & Low                                                                        & 95.1          & 77.5          & 98.5          & 0.33          & 42             \\
31           & Good                                                                      & Moderate                                                                   & 98.8          & 99.2          & 100           & 0.31          & 40             \\ 
\bottomrule
\textbf{Avg} &                                                                           &                                                                            & \textbf{93.6} & \textbf{76.3} & \textbf{96.4} & \textbf{0.37} & \textbf{42.6}  \\
\bottomrule
\end{tabular}}
\caption{Results of our method on CAT08 test set. For each test case, overlap (OV, in \%), overlap until first error (OF, in \%) and clinically relevant overlap (OT, in \%) ,average accuracy inside (AI, in mm), time taken for coronary tree extraction (T, in s)  along with subjective image quality and calcium score is shown.}
\label{table: table2}
\end{table}

Table \ref{table: table2} shows the performance of the algorithm on the testing set of MICCAI 2008 challenge. An average overlap of \textbf{93.6\%}, clinically relevant overlap of \textbf{96.4\%} and overlap until first error of \textbf{76.3\%} was obtained for these 24 CCTA scans. Cases 8, 10 and 27 required one additional seed point due to failure in the detection of bifurcations for one of the vessels. There are significant motion artifacts present case 26 which hamper the bifurcation detection. Hence, additional seed points are provided for 3 of the vessels in this CCTA image. There are 128 vessels present in the training and test set of CAT08 dataset. Overall, 125 vessels were automatically detected without requiring any seed points.

\begin{table}[t!]
\centering
\resizebox{\columnwidth}{!}{
  \begin{tabular}{l|r|r|r|r|r} 
  \toprule
  \textbf{Method}  & \textbf{OV}   & \textbf{OF}   & \textbf{OT }  & \textbf{AI}  & \textbf{T} \\ 
  \hline
  \textcolor[rgb]{0.7,0,0}{AuCoTrack } & 93.6 & 76.3 & \textbf{96.4} & 0.37 & \textbf{42}  \\
  Zheng et al.    &         93.5  & \textbf{76.5} & 95.6 & \textbf{0.20}&  60 \\
  Kitamura et al. &         90.6  &         70.9  & 92.5 &         0.25 & 160 \\
  Yang et al.     & \textbf{93.7} &         74.2  & 95.9 &         0.30 & 120 \\ 
  \hdashline
  \textcolor[rgb]{0,0,0}{Wolterink et al. (Interactive)} & 93.7  & 81.5 & 97.0 & 0.21 & 10 \\
  \bottomrule
  \end{tabular}
}
\caption{The comparison of our proposed AuCoTrack algorithm and the top automatic coronary artery centerline extraction techniques in terms of overlap (OV, in \%), overlap until first error (OF, in \%) and clinically relevant overlap (OT, in \%), average accuracy inside (AI, in mm) and time taken (T, in s). The interactive CNN-based method by \cite{WOLTERINK201946} is separated by a dotted line.}
\label{table: table3}
\end{table}
Table \ref{table: table3} shows the comparison of the  performance of our algorithm \textbf{AuCoTrack} against the current automatic coronary centerline extraction techniques and the state-of-the-art CNN-based technique which requires at-least one point per vessel for the centerline extraction. Our proposed method achieves better overlap rank of \textbf{9.87} than other automatic techniques of Zheng \textbf{10.43}, Kitamura \textbf{13.81} and Yang \textbf{10.55}. Wolterink's CNN-based approach requires at least one seed point per vessels. AuCoTrack successfully detected  approximately 95\% percent of the vessels in CAT08 dataset requiring no user interaction. However, Wolterink's approach gives OV, OF and OT values of 93.7\%, 81.5\%  and 96\% respectively. Our method achieves almost the same performance as Wolterink's CNN approach while reducing the need of interaction to almost zero. Our approach requires an average runtime of \textbf{42.6 s} using a GTX 1080 GPU to extract the entire coronary tree. Our method is faster than all other automatic approaches, but the comparison is difficult as the computational resources of all the methods are not same.

\begin{figure*}[t!]
\centerline{\includegraphics[width=0.9\textwidth]{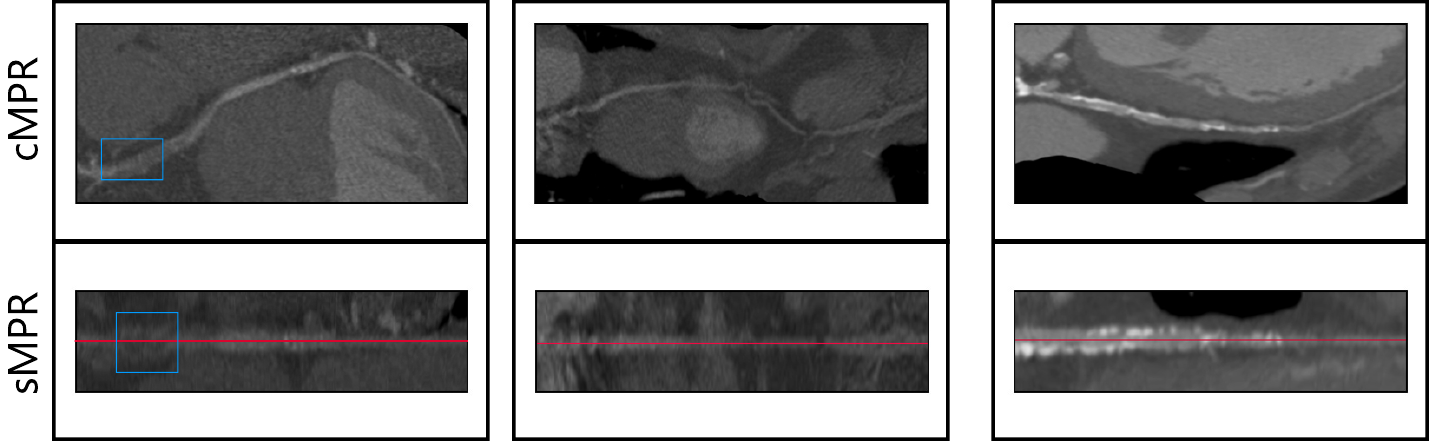}}
\caption{Stretched multiplanar reformation (sMPR) and curved multiplanar reformation (cMPR) images constructed from extracted centerlines using  our proposed method \textbf{``AuCoTrack''}. The blue box in case \#0 from training set of CAT08 challenge (left) shows the presence of severe stenosis. Case \#26 from the testing set of the CAT08 challenge (middle) has severe motion artifacts. MPR and cMPR images of a case from the in-house dataset (right) show the presence of coronary calcification. The red lines show the extracted centerline overlayed on the sMPR image.}
\label{figure: cprmpr}
\end{figure*}

\section{Discussion}
\label{sec:discussion}
We aimed to provide a deep learning-based automatic approach for centerline extraction in CCTA images. The proposed algorithm was first tested on an in-house dataset containing CCTA scans acquired at multiple sites. The sweeps for hyper-parameter tuning were performed on in-house dataset using 33 CCTA scans for training and 10 CCTA scans for validation. The method was then evaluated using four-fold cross validation on these CCTA scans. A high average clinically relevant overlap of 89.1\% and average sensitivity of 87.1\%  was obtained. The average accuracy inside for the in-house dataset was reported to 0.34 mm which is less than the average voxel dimensions. 

The generalization of this approach was then evaluated by testing the proposed algorithm on the CAT08 training dataset. The model from four-fold cross validation on in-house dataset was used to extract centerlines for CAT08 training dataset. The images from CAT08 training dataset contained considerable variability in terms of calcium scores and image quality. An average overlap of 93.4\% and  clinically relevant overlap of 95.9\% was obtained on evaluating CAT08 training dataset as a test set. 

In order to benchmark the performance, we also tested the algorithm by submitting the extracted centerlines of 96 vessels on the evaluation framework of CAT08 challenge. In order to extract centerlines for CAT08 test set, the method was trained on all 43 CCTA scans from the in-house dataset and 8 CCTA scans from CAT08 training set. The bifurcation detection failed in 7 out of these 128 vessels. A seed point was required in these cases in order to retrieve these coronary arteries.

The proposed algorithm achieves better overlap rank than the previously available fully automatic coronary artery centerline extraction algorithms. An overlap rank of 9.87 was achieved by AuCoTrack while the top three automatic algorithms on the CAT08 leaderboard by \cite{Zheng2013}, \cite{Kitamura2012} and \cite{Yang2011} had an overlap rank of 10.43, 13.81 and 10.55 respectively. The accuracy inside for the CAT08 testing data set was 0.37 mm. The state-of-the-art automatic centerline extraction algorithm by \cite{Zheng2013} utilizes segmentation masks for their model driven and data driven approach. Their algorithm uses 108 CCTA scans from their proprietary dataset. Our proposed algorithm achieves state-of-the-art overlap metrics on training set of CAT08 challenge when trained with only 33 CCTA scans from a distinct in-house dataset. Hence, this method can be trained efficiently on low number of CCTA scans.

Table \ref{table: table2} shows the evaluation metrics for the testing set of CAT08 challenge with variable image quality and calcium score. An average overlap of 87.8\%, 95.4\% and 94.4\% was achieved for CCTA scans with poor, moderate and good image quality respectively. Poor image quality is defined by the presence of image degrading artifacts and evaluation is only possible with low confidence \cite{Schaap2009701}. Our algorithm's performance is effected by poor image quality but the performance is consistent over moderate and good image quality. CCTA images with calcium scores of low, moderate and severe had an average overlap of 92.9\%, 94.0\% and 94.2\% respectively. This shows the performance of the algorithm is not strongly effected by the presence of coronary calcification.

The proposed algorithm aims at extracting the entire coronary tree from a single seed point. The comparative low metrics for the CCTA scans in in-house dataset as compared to CAT08 dataset can also be attributed to the fact that average number of annotated arteries in the in-house dataset is 9 as compared to 4 in the CAT08 dataset. High clinically relevant overlap in the cases with large number of annotated arteries show that our algorithm is capable of extracting the entire coronary tree. Some of the arteries that may be missed can retrieved by a single seed point. 
 
\cite{WOLTERINK201946} achieved near state-of-the-art performance as an interactive method for the CAT08 dataset. This method was based on a CNN classifier which simultaneously predicts direction to the centerlines and radius. The main constraint of this method is that it requires one or more seed points per vessel. Our method removes the requirement of seed points per vessel. A seed point is required only when the bifurcation detection fails for the corresponding artery. Successful vessel detection was observed in 95 percent of the cases in CAT08 dataset. The ostia points required for tracker initialization were automatically obtained from an MBS model. \cite{WOLTERINK201946}'s tracker termination is guided by a moving average entropy criteria which fails in case of a severe stenosis. This is the reason that in some of the cases more than one point is required per vessel in order to warm-start the tracking process. Our method utilizes a model trained on patches beyond the end point in order to determine if the end of a coronary artery has been reached. We employ a voting mechanism of stop patch classification as well as moving average entropy in order to terminate the tracking. 

The number of annotated centerlines in the in-house dataset varies from 4 to 20. This attributes to severe label noise for bifurcation classification because the underrepresented bifurcation patches may be labeled as normal patches in cases where the number of annotated centerlines are low. This problem may be mitigated by labeling the missing bifurcation points on the annotated arteries. Alternatively, active learning or label noise suppression strategies can also be explored for the solution \cite{Karimi2020, Wang2008}.

The extraction takes on average 42.1 s for the entire coronary tree in CAT08 dataset. The time complexity of the tracking algorithm in Listing \ref{listing : lst1} in worst case is essentially bounded by $O(n)$ where $n$ represents the total number of tracked points. Hence, the total time taken depends on the size of the extracted coronary tree. This time can be reduced by many folds by optimizing the tracker. The different sub-trees can be processed in a parallel fashion making use of the bifurcation predication. Within the same sub-tree, different threads can access the main active queue and process the data in a parallel fashion keeping track of the visited points. The tracking result on the CCTA image can be displayed in real time as tracking is being performed.

Figure \ref{figure: cprmpr} shows stretched multiplanar reformation (sMPR) and curved multiplanar reformation (cMPR) images reconstructed using centerlines obtained from \textbf{``AuCoTrack''} algorithm. These reformatted images can be directly used for the diagnosis of coronary artery disease. The proposed algorithm makes use of the local intensity information in the patches in order detect the direction to the centerlines and the bifurcation. We have shown that the model trained on CCTA images from the in-house dataset works well for the images from old Siemens scanners for CAT08 dataset.

Overall, AuCoTrack's runtime (non-optimised version takes 40s per case) combined with its high overlap performance are remarkable given its simplicity. The generalization is demonstrated by the fact that it was trained on an in-house dataset and tested on the CAT08 dataset with considerable variability. Since the algorithm is based on local intensity values of the patches, the same proposed pipeline/model can be used to obtain centerlines in other applications e.g. rib centerline extraction.

Accuracy can be further boosted by a re-centering of the extracted centerlines. Further seed points for potentially missed vessels can be obtained by a rough pre-segmentation scheme. False positive detections of vessels e.g. coronary veins or thoracic arteries can be reduced by combing further anatomical prior information with the local tracker. Initialisation at intermediate landmarks different from the ostiae is feasible and can be further explored. The algorithm can be parallelized much stronger in order to decrease runtime per case. Also, multiscale processing is a possible option for future work. The tracking scheme is very generic and versatile and can also be used to track other 3D tubular structures such as airways, rib centerlines and other blood vessels.

\bibliographystyle{acm}
\bibliography{preprint}

\begin{thebibliography}{10}

\bibitem{Cademartiri2007129}
{\sc Cademartiri, F., Grutta, L., Palumbo, A., Malagutti, P., Pugliese, F.,
  Meijboom, W., Baks, T., Mollet, N., Bruining, N., Hamers, R., and Feyter, P.}
\newblock Non-invasive visualization of coronary atherosclerosis: State-of-art.
\newblock {\em Journal of Cardiovascular Medicine (Hagerstown, Md.) 8\/} (04
  2007), 129--37.

\bibitem{Cetin2012}
{\sc Cetin~Karayumak, S., Demir, A., Yezzi, A., Degertekin, M., and Unal, G.}
\newblock Vessel tractography using an intensity based tensor model with branch
  detection.
\newblock {\em IEEE transactions on medical imaging 32\/} (11 2012).

\bibitem{Cetin2015}
{\sc Cetin~Karayumak, S., and Unal, G.}
\newblock A higher-order tensor vessel tractography for segmentation of
  vascular structures.
\newblock {\em IEEE transactions on medical imaging 34\/} (04 2015).

\bibitem{Ecabert2008}
{\sc Ecabert, O., Peters, J., Schramm, H., Lorenz, C., Berg, J., Walker, M.,
  Vembar, M., Olszewski, M., Subramanyan, K., Lavi, G., and Weese, J.}
\newblock Automatic model-based segmentation of the heart in {CT} images.
\newblock {\em IEEE transactions on medical imaging 27\/} (10 2008), 1189--201.

\bibitem{Frangi1998}
{\sc Frangi, A.~F., Niessen, W.~J., Vincken, K.~L., and Viergever, M.~A.}
\newblock Multiscale vessel enhancement filtering.
\newblock In {\em Medical Image Computing and Computer-Assisted Intervention\/}
  (Berlin, Heidelberg, 1998), W.~M. Wells, A.~Colchester, and S.~Delp, Eds.,
  Springer Berlin Heidelberg, pp.~130--137.

\bibitem{Frangi2000}
{\sc Frangi, R., Niessen, W., Vincken, K., and Viergever, M.}
\newblock Multiscale vessel enhancement filtering.
\newblock {\em Med. Image Comput. Comput. Assist. Interv. 1496\/} (02 2000).

\bibitem{Friman2020}
{\sc Friman, O., Kuehnel, C., and Peitgen, H.-O.}
\newblock Coronary centerline extraction using multiple hypothesis tracking and
  minimal paths.

\bibitem{Hampe20196}
{\sc Hampe, N., Wolterink, J., Velzen, S., Leiner, T., and Išgum, I.}
\newblock Machine learning for assessment of coronary artery disease in cardiac
  {CT}: A survey.
\newblock {\em Frontiers in Cardiovascular Medicine 6\/} (11 2019).

\bibitem{Huang2018608}
{\sc Huang, W., Huang, L., Lin, Z., Huang, S., Chi, Y., Zhou, J., Zhang, J.-M.,
  Tan, R.~S., and Zhong, L.}
\newblock Coronary artery segmentation by deep learning neural networks on
  computed tomographic coronary angiographic images.
\newblock vol.~2018, pp.~608--611.

\bibitem{Kamnitsas2016}
{\sc Kamnitsas, K., Ledig, C., Newcombe, V., Simpson, J., Kane, A., Menon, D.,
  Rueckert, D., and Glocker, B.}
\newblock Efficient multi-scale {3D CNN} with fully connected {CRF} for
  accurate brain lesion segmentation.
\newblock {\em Medical Image Analysis 36\/} (03 2016).

\bibitem{Karimi2020}
{\sc Karimi, D., Dou, H., Warfield, S., and Gholipour, A.}
\newblock Deep learning with noisy labels: Exploring techniques and remedies in
  medical image analysis.
\newblock {\em Medical Image Analysis 65\/} (06 2020), 101759.

\bibitem{Keinert2015}
{\sc Keinert, B., Innmann, M., Sänger, M., and Stamminger, M.}
\newblock Spherical {F}ibonacci mapping.
\newblock {\em ACM Transactions on Graphics 34\/} (10 2015), 1--7.

\bibitem{Kitamura2012}
{\sc Kitamura, Y., Li, Y., and Ito, W.}
\newblock Automatic coronary extraction by supervised detection and shape
  matching.
\newblock {\em Proceedings - International Symposium on Biomedical Imaging\/}
  (05 2012), 234--237.

\bibitem{Krissian2018}
{\sc Krissian, K., Bogunović, H., Pozo, J., Villa-Uriol, M.-C., and Frangi,
  A.}
\newblock Minimally interactive knowledge-based coronary tracking in {CTA}
  using a minimal cost path.

\bibitem{Malakar2016}
{\sc Malakar, A., Choudhury, D., Halder, B., Paul, P., Uddin, D.~A., and
  Chakraborty, S.}
\newblock A review on coronary artery disease, its risk factors, and
  therapeutics.
\newblock {\em Journal of Cellular Physiology 234\/} (02 2019).

\bibitem{Peach201111}
{\sc Paech, D., and Weston, A.}
\newblock A systematic review of the clinical effectiveness of 64-slice or
  higher computed tomography angiography as an alternative to invasive coronary
  angiography in the investigation of suspected coronary artery disease.
\newblock {\em BMC Cardiovascular Disorders 11\/} (06 2011), 32.

\bibitem{salahuddin20aucotrack}
{\sc Salahuddin, Z., Lenga, M., and Nickisch, H.}
\newblock Multi-resolution 3d convolutional neural networks for automatic
  coronary centerline extraction in cardiac {CT} angiography scans.
\newblock In {\em Master Thesis Proceedings 2018-20\/} (2020), Erasmus Mundus
  Joint Master Degree in Medical Imaging and Applications (MAIA),
  \url{http://eia.udg.edu/~aoliver/maiaDocs/bookMaia3rd_small.pdf},
  pp.~207--222.

\bibitem{Schaap2009701}
{\sc Schaap, M., Metz, C., Walsum, T., van~der Giessen, A., Weustink, A.,
  Mollet, N., Bauer, C., Bogunović, H., Castro, C., Deng, X., Dikici, E.,
  O'Donnell, T., Frenay, M., Friman, O., Hernandez~Hoyos, M., Kitslaar, P.,
  Krissian, K., Kühnel, C., Luengo-Oroz, M., and Niessen, W.}
\newblock Standardized evaluation methodology and reference database for
  evaluating coronary artery centerline extraction algorithms.
\newblock {\em Medical image analysis 13\/} (07 2009), 701--14.

\bibitem{Stimpel2018}
{\sc Stimpel, B., Wetzl, J., Forman, C., Schmidt, M., Maier, A., and Unberath,
  M.}
\newblock Automated curved and multiplanar reformation for screening of the
  proximal coronary arteries in {MR} angiography.
\newblock {\em Journal of Imaging 4\/} (10 2018), 124.

\bibitem{Tavakol2012}
{\sc Tavakol, M., Ashraf, S., and Brener, S.}
\newblock Risks and complications of coronary angiography: A comprehensive
  review.
\newblock {\em Global journal of health science 4\/} (09 2012), 65--93.

\bibitem{Wang2008}
{\sc Wang, C., and Smedby, O.}
\newblock An automatic seeding method for coronary artery segmentation and
  skeletonization in {CTA}.

\bibitem{WHO2018}
{\sc WHO}.
\newblock Global health estimates 2016: Deaths by cause, age, sex, by country
  and by region, 2000-2016., 2018.

\bibitem{WOLTERINK201946}
{\sc Wolterink, J.~M., {van Hamersvelt}, R.~W., Viergever, M.~A., Leiner, T.,
  and Išgum, I.}
\newblock Coronary artery centerline extraction in cardiac {CT} angiography
  using a {CNN}-based orientation classifier.
\newblock {\em Medical Image Analysis 51\/} (2019), 46 -- 60.

\bibitem{Yang2011}
{\sc Yang, G., Kitslaar, P., Frenay, M., Broersen, A., Boogers, M., Bax, J.,
  Reiber, J., and Dijkstra, J.}
\newblock Automatic centerline extraction of coronary arteries in coronary
  computed tomographic angiography.
\newblock {\em The International Journal of Cardiovascular Imaging 28\/} (06
  2011), 921--33.

\bibitem{Zheng2013}
{\sc Zheng, Y., Tek, H., and Funka-Lea, G.}
\newblock Robust and accurate coronary artery centerline extraction in {CTA} by
  combining model-driven and data-driven approaches.
\newblock vol.~16, pp.~74--81.

\bibitem{Zreik2019}
{\sc {Zreik}, M., {van Hamersvelt}, R.~W., {Wolterink}, J.~M., {Leiner}, T.,
  {Viergever}, M.~A., and {Išgum}, I.}
\newblock A recurrent {CNN} for automatic detection and classification of
  coronary artery plaque and stenosis in coronary {CT} angiography.
\newblock {\em IEEE Transactions on Medical Imaging 38}, 7 (2019), 1588--1598.

\end{thebibliography}




\end{document}